\documentclass{LMCS}

\usepackage{enumerate}
\usepackage{hyperref}
\usepackage{graphicx}

\def\cls#1{\overline{#1}} 
\def\st{\bigm|}
\def\dto{\Rightarrow} 
\def\implies{\rightarrow}
\def\yields{\vdash}
\def\D{\mathcal{D}}
\def\U{\mathcal{U}}
\def\B{\mathcal{B}}

\def\R{\mathcal{R}}
\def\myB{{\B}^{\star}_{\gamma}}

\long\def\marginal#1{}
\long\def\future#1{}

\def\doi{6 (2:3) 2010}
\lmcsheading%
{\doi}
{1--33}
{}
{}
{Jan.~13, 2009}
{Jun.~22, 2010}
{}   

\begin{document}

\title[Redundancy and Bases for Association Rules]{Redundancy, 
Deduction Schemes, and 
Minimum-Size Bases for Association Rules}


\author{Jos\'e L. Balc\'azar} 
\address{Dep.~de Matem\'aticas, Estad{\'\i}stica y Computaci{\'o}n\\
Universidad de Cantabria\\
Santander, Spain}
\email{joseluis.balcazar@unican.es}
\thanks{This work is supported in part by project
TIN2007-66523 (FORMALISM)
of Programa Nacional de Investigaci\'on,
Ministerio de Ciencia e Innovaci\'on (MICINN), Spain, and by the 
PASCAL2 Network of Excellence of the European Union.}




\begin{abstract}
\noindent 
Association rules are among the 
most widely employed data analysis
methods in the field of Data Mining. 
An association rule is a form of 
partial implication between two
sets of binary variables. In the most 
common approach, association rules are 
parametrized by a 
lower bound on their confidence, 
which is the empirical conditional 
probability of their consequent 
given the antecedent, and/or by 
some other parameter bounds such as 
``support'' or deviation from 
independence. We study here notions 
of redundancy among association rules 
from a fundamental perspective. 
We see each transaction in a 
dataset as an interpretation 
(or model) in the propositional 
logic sense, and consider existing
notions of redundancy, that is, 
of logical entailment, among association
rules, of the form 
``any dataset in which this first rule holds must obey 
also that second rule, therefore the second is redundant''.
We discuss several existing 
alternative definitions of redundancy 
between association rules 
and provide new characterizations 
and relationships among them. 
We show that the main alternatives 
we discuss correspond actually to just 
two variants, which differ in the 
treatment of full-confidence implications. 
For each of these two notions of redundancy, we 
provide a sound and complete deduction calculus, 
and we show how to construct complete bases 
(that~is, axiomatizations) of absolutely minimum size 
in terms of the number of rules. 
We explore finally an approach to
redundancy with respect to several
association rules, and fully characterize its
simplest case of two partial premises.
\end{abstract}

\maketitle

\section{Introduction}

The relatively recent discipline of Data Mining 
involves a wide spectrum of techniques, inherited
from different origins such as Statistics, Databases,
or Machine Learning. Among them, Association Rule Mining 
is a prominent conceptual tool and, possibly, a cornerstone 
notion of the field, if there is one. Currently, 
the amount of available knowledge regarding association rules
has grown to the extent that the tasks of creating complete 
surveys and websites that maintain pointers to related
literature become daunting. A survey, with plenty of
references, is~\cite{CegRod}, and additional materials 
are available in \cite{HahslerWeb}; see also 
\cite{AIS}, 
\cite{AMSTV},
\cite{Freitas},
\cite{PasBas}, 
\cite{Zaki}, 
\cite{ZO}, 
and the references and discussions in their introductory sections.

Given an agreed general set of ``items'', 
association rules are defined with respect to a 
dataset that consists of ``transactions'', each 
of which is, essentially, a set of items. 
Association rules are customarily written as $X\to Y$, 
for sets of items 
$X$ and $Y$, and they hold in the given dataset with a 
specific ``confidence'' quantifying how often $Y$ appears 
among the transactions in which $X$ appears.

A close relative of the notion of association rule, namely, 
that of exact implication in the standard propositional
logic framework, or, equivalently, association rule that 
holds in 100\% of the cases, has been studied in 
several guises. 
Exact implications are equivalent to conjunctions 
of definite Horn clauses: 
the fact, well-known in logic and knowledge representation, 
that Horn theories are exactly those closed under 
bitwise intersection of propositional models 
leads to a strong connection with Closure Spaces,
which are characterized by closure under intersection
(see~the discussions in \cite{DP}~or~\cite{KR}).
Implications are also very closely related to 
functional dependencies in databases. Indeed,
implications, as well as functional dependencies, 
enjoy analogous, clear, robust, hardly disputable notions 
of redundancy that can be defined equivalently both 
in semantic terms and through the same syntactic calculus. 
Specifically, for the semantic notion of entailment, 
an implication $X\to Y$ is entailed from a set 
of implications $\R$ if every dataset in which all
the implications of $\R$ hold must also satisfy $X\to Y$;
and, syntactically, it is known that this happens if and 
only if $X\to Y$ is derivable from $\R$ via the Armstrong
axiom schemes, namely, 
Reflexivity ($X\to Y$ for $Y\subseteq X$), 
Augmentation (if $X\to Y$ and $X'\to Y'$ then $XX'\to YY'$,
where juxtaposition denotes union) and 
Transitivity (if $X\to Y$ and $Y\to Z$ then $X\to Z$).

Also, such studies have provided a number of ways to
find implications (or functional dependencies) that
hold in a given dataset, and to
construct small subsets of a large set of implications, 
or of functional dependencies, from which the whole set
can be derived; in Closure Spaces and in Data Mining 
these small sets are usually called ``bases'', whereas
in Dependency Theory they are called ``covers'', and they
are closely related to deep topics such as hypergraph theory.
Associated natural notions of minimality (when no
implication can be removed), minimum size, and 
canonicity of a cover or basis do exist; 
again it is inappropriate to try
to give a complete set of references here, but see, 
for instance, 
\cite{DP},
\cite{EiterG},
\cite{GW}, 
\cite{GD}, 
\cite{GunoEtAl},
\cite{KR},
\cite{PT},
\cite{Wild},
\cite{ZO},
and the references therein.

However, the fact has been long acknowledged (e.g.~already 
in \cite{Lux}) that, often, it is inappropriate to search 
only for absolute implications in the analysis of real world 
datasets. Partial rules are defined in 
relation to their ``confidence'': 
for a given rule $X\to Y$, the ratio of 
how often $X$ and $Y$ are seen~together to how often $X$ 
is seen. Many other alternative 
measures of intensity of implication exist
\cite{Garriga}, \cite{GH}; 
we keep our focus on confidence because, besides
being among the most common ones, 
it has a natural interpretation for educated users 
through its correspondence with the observed conditional 
probability.

The idea of 
restricting the exploration for association rules to frequent 
itemsets, with respect to a support threshold, gave rise 
to the most widely discussed and applied algorithm, called
Apriori \cite{AMSTV}, and to an intense research activity. 
Already with full-confidence implications, the 
output of an association mining process often consists of
large sets of rules, and a well-known difficulty in applied 
association rule mining lies in that, on large datasets, and 
for sensible settings of the confidence and support thresholds
and other parameters, huge amounts of association rules are often 
obtained.
Therefore, besides the interesting progress in the topic of how to
organize and~query the rules discovered (see
\cite{LiuHsuMa},
\cite{LiuHuHsu},
\cite{TuLiu}),
one research topic that has been worthy of attention 
is the identification of patterns that indicate redundancy 
of rules, and ways to avoid that redundancy;
and each proposed notion of redundancy opens up
a major research problem, namely, to provide
a general method for constructing bases of 
minimum size with respect to that notion of 
redundancy.

For partial rules, the Armstrong schemes 
are not valid anymore. Reflexivity 
does hold, 
but
Transitivity takes a different form that affects 
the confidence of the rules: if the rule $A\to B$ 
(or $A\to AB$, which is equivalent) and the rule 
$B\to C$ both hold with confidence at least $\gamma$, 
we still know nothing about the confidence 
of $A\to C$; even the fact that both $A\to AB$ 
and $AB\to C$ hold with confidence at least $\gamma$ 
only gives us a confidence lower bound of 
$\gamma^2<\gamma$ for $A\to C$ (assuming~\hbox{$\gamma<1$}). 
Augmentation does not hold at all; indeed,
enlarging the antecedent of a rule of confidence at least
$\gamma$ may give a rule with much smaller confidence, 
even zero: think of a case where most of the 
times $X$ appears it comes with~$Z$, but 
it only comes with $Y$ when $Z$ is not present; then 
the confidence of $X\to Z$ may be high whereas the 
confidence of $XY\to Z$ may be null. Similarly, 
if the confidence of $X\to YZ$ 
is high, it means that $Y$ and $Z$ 
appear together in most of the transactions
having $X$, whence the confidences of $X\to Y$
and $X\to Z$ are also high; but, with respect to the converse,
the fact that both $Y$ and $Z$ appear in fractions 
at least $\gamma$ of the transactions having $X$
does not inform us that they show up {\em together} 
at a similar ratio of these transactions: only a ratio 
of $2\gamma-1<\gamma$ is guaranteed as a lower bound.
In fact,
if we look only for association rules with singletons as
consequents (as in some of the analyses in \cite{AgYu},
or in the ``basic association rules'' of \cite{LiHa}, or
even in the traditional approach to association rules \cite{AIS}
and the useful {\tt apriori} implementation of 
Borgelt available on the web~\cite{BorgeltApriori}) 
we are almost certain to 
lose information. 
As a consequence of these failures of the Armstrong schemes,
the canonical and minimum-size cover construction methods
available for implications or functional dependencies are
not appropriate for partial association rules.

On the semantic side,
a number of formalizations of the intuition of redundancy
among association rules exist in the literature,
often with proposals for defining irredundant bases
(see
\cite{AgYu}, 
\cite{CrisSim},
\cite{KryszPAKDD}, 
\cite{Lux}, 
\cite{PasBas},
\cite{PhanLuongICDM},~\cite{Zaki}, 
the~survey~\cite{Krysz}, and
section~6 of the survey~\cite{CegRod}). 
All of these are weaker than the notion that we would
consider natural by comparison with implications
(of which we start the study in the last section
of this paper). 
We observe here that 
one may wish to fulfill two different roles
with a basis, and that both appear (somewhat mixed)
in the literature: as a computer-supported 
data structure from which confidences and 
supports of rules are computed (a role for 
which we use the closures lattice instead) 
or, in our choice, as a means of providing 
the user with a smallish set of association rules 
for examination and, if convenient, posterior 
enumeration of the rules that follow from 
each rule in the basis.
That is, we will not assume 
to have available, nor to wish to compute, exact values 
for the confidence, but only discern whether it stays 
above a certain user-defined threshold.
We compute actual confidences out of the
closure lattice only at the time of writing 
out rules for the user.

This paper focuses mainly on several such notions of redundancy, 
defined in a rather 
general way, by resorting to confidence and support 
inequalities: essentially, a rule is redundant with respect 
to another if it has at least the same confidence and 
support of the latter {\em for every dataset}.
We also discuss variants of this proposal and other existing 
definitions given in set-theoretic terms. 
For the most basic notion of
redundancy, we provide formal proofs of the
so far unstated 
equivalence among several published 
proposals, including a syntactic calculus and a formal proof
of the fact, also previously unknown, that the existing basis 
known as the Essential Rules or the Representative Rules
(\cite{AgYu}, \cite{KryszPAKDD}, \cite{PhanLuongICDM}) 
is of absolutely minimum size.

It is natural to wish further progress 
in reducing the size of the basis. Our theorems 
indicate that, in order to reduce further 
the size without losing information, more powerful 
notions or redundancy must be deployed. We consider 
for this role the proposal of 
handling separately, 
to a given extent, full-confidence implications from 
lower-than-1-confidence rules, in order to profit 
from their very different combinatorics. This separation
is present in many constructions of bases for
association rules \cite{Lux}, \cite{PasBas}, \cite{Zaki}.
We discuss corresponding 
notions of redundancy and completeness, and prove 
new properties of these notions; we give a
sound and complete deductive calculus for this 
redundancy; and we refine the existing basis constructions 
up to a point where we can prove again that we attain the 
limit of the redundancy notion. 

Next, we discuss yet another potential for
strengthening the notion of redundancy. So far,
all the notions have just related one partial
rule to another, possibly in the presence of
full implications. Is it possible to combine
two partial rules, of confidence at least $\gamma$,
and still obtain a partial rule obeying that
confidence level? Whereas the intuition is that
these confidences will combine together to yield
a confidence lower than $\gamma$, we prove that
there is a specific case where a rule of 
confidence at least $\gamma$ is nontrivially entailed 
by two of them. We fully characterize this case
and obtain from the caracterization yet another
deduction scheme. We hope that further progress
along the notion of a {\em set} of partial rules
entailing a partial rule will be made along the
coming years.

Preliminary versions of the results in sections
\ref{dedplain},
\ref{redundcalculus},
\ref{clocalcsoundcompl},
and \ref{closbasedent} 
have been presented~at Discovery Science 2008 \cite{Bal08b};
preliminary versions of the remaining results
(except those in section~\ref{suppbound}, 
which are newer and unpublished)
have been presented at ECMLPKDD~2008~\cite{Bal08}.

\section{Preliminaries}

Our notation and terminology are quite standard
in the Data Mining literature. All our developments 
take place in the presence of a ``universe'' set $\U$ 
of atomic elements called {\em items}; their absence
or presence in sets or items plays the 
same role as binary-valued attributes of a relational table. 
Subsets of $\U$ are called {\em itemsets}. A dataset $\D$ 
is assumed to be given; it consists of 
transactions, each of which is an itemset 
labeled by a unique transaction identifier.
The identifiers allow us to distinguish among 
transactions even if they share the same itemset. 
Upper-case, often subscripted letters from the end 
of the alphabet, like $X_1$~or~$Y_0$, 
denote itemsets. Juxtaposition denotes 
union of itemsets, as in $XY$; and 
$Z\subset X$ denotes {\em proper} subsets,
whereas $Z\subseteq X$ is used for the
usual subset relationship with potential 
equality. 

For a transaction $t$, we denote $t\models X$ the fact 
that $X$ is a subset of the itemset corresponding 
to $t$, that is, the transaction satisfies the
minterm corresponding to $X$ in the propositional
logic sense.

{}From the given dataset we obtain a notion of support
of an itemset: $s_{\D}(X)$ is the cardinality of the 
set of transactions that include it, $\{ t\in\D\st t\models X\}$;  
sometimes, abusing language slightly, we also refer to that set of 
transactions itself as support. 
Whenever $\D$ is clear,
we drop the subindex: $s(X)$. 
Observe that $s(X)\geq s(Y)$ 
whenever $X\subseteq Y$;
this is immediate from the 
definition. Note that many
references resort to a normalized
notion of support by dividing
by the dataset size. We chose not to,
but there is no essential issue here.
Often, research work in 
Data Mining
assumes that a threshold on the support has been provided
and that only sets whose support is above the threshold
(then called ``frequent'') are to be considered. We will
require this additional constraint occassionally for the
sake of discussing the applicability of our developments.

We immediately obtain by standard means 
(see, for instance, \cite{GW} or \cite{Zaki})~a 
notion of closed itemsets, namely, 
those that cannot be enlarged while 
maintaining the same support. 
The function 
that maps each itemset to
the smallest closed set that contains it is known
to be monotonic, extensive, and idempotent,
that is, it is a closure operator.
This notion will be reviewed in more detail
later~on. Closed sets whose support is above
the support threshold, if given, are usually
termed closed frequent sets.

Association rules are pairs of itemsets, denoted 
as $X\to Y$ for itemsets $X$ and $Y$. Intuitively,
they suggest the fact that $Y$ occurs particularly 
often among the transactions in which $X$ occurs. 
More precisely, each such rule has a confidence 
associated: the confidence $c_{\D}(X\to Y)$ of 
an association rule $X\to Y$ in a dataset $\D$ is 
$\frac{s(XY)}{s(X)}$.
As with support, often we drop the subindex $\D$. 
The support in $\D$ of the association rule
$X\to Y$ is $s_{\D}(X\to Y) = s_{\D}(XY)$.

We can switch rather freely between right-hand sides
that include the left-hand side and right-hand sides that don't: 

\begin{defi}
Rules $X_0\to Y_0$ and $X_1\to Y_1$ are 
{\em equivalent by reflexivity}
if $X_0 = X_1$ and $X_0Y_0 = X_1Y_1$.
\end{defi}

Clearly, $c_{\D}(X\to Y) = c_{\D}(X\to XY) = c_{\D}(X\to X'Y)$
and, likewise, $s_{\D}(X\to Y) = s_{\D}(X\to XY) = s_{\D}(X\to X'Y)$
for any $X'\subseteq X$; that is, the support and
confidence of rules that are equivalent by
reflexivity always coincide.
A minor notational issue that we must point out is that,
in some references, the left-hand side of a rule is required 
to be a subset of the right-hand side,
as in \cite{Lux} or \cite{PhanLuongICDM},
whereas many others require the left- and right-hand
sides of an association rule to be disjoint, such
as \cite{Krysz} or the original~\cite{AIS}. 
Both the rules whose left-hand side is a subset
of the right-hand side, and the rules that have
disjoint sides, may act as canonical representatives
for the rules equivalent to them by reflexivity.
We state explicitly one version of this immediate 
fact for later reference:

\begin{prop} 
\label{disjointness}
If rules $X_0\to Y_0$ and $X_1\to Y_1$ 
are equivalent by reflexivity, 
$X_0\cap Y_0 = \emptyset$, 
and $X_1\cap Y_1 = \emptyset$, then 
they are the same rule:
$X_0 = X_1$ and $Y_0 = Y_1$.
\end{prop}

In general, we do allow, along our 
development, rules where the left-hand side, 
or a part of it, appears also at the 
right-hand side, because by doing so we will be able
to simplify the mathematical arguments.
We will assume here that, at the
time of printing out the rules found, that is, 
for user-oriented output, the items
in the left-hand side are removed from the
right-hand side;
accordingly, we write our
rules 
sometimes as $X\to Y-X$
to recall this convention.

Also, many references require the right-hand side of
an association rule to be nonempty, or even both sides. 
However, empty sets can be handled with no difficulty 
and do give meaningful, albeit uninteresting, rules. 
A partial rule $X\to\emptyset$ with an empty right-hand 
side is equivalent by reflexivity to $X\to X$, or to 
$X\to X'$ for any $X'\subseteq X$, and all of these rules
have always confidence~1. A partial rule with empty left-hand 
side, as employed, for instance, in \cite{Krysz},
actually gives the normalized support of the right-hand 
side as confidence value:

\begin{fact}
In a dataset $\D$ of $n$ transactions,
$c(\emptyset\to Y) = s(Y)/n$.
\label{emptysides}
\end{fact}

Again, these sorts of rules could 
be omitted from user-oriented output, but considering them 
conceptually valid simplifies the mathematical development.
We also resort to the convention that, if $s(X)=0$ 
(which implies that $s(XY)=0$ as well) we redefine the 
undefined confidence $c(X\to Y)$ as 1, since the intuitive
expression ``all transactions having~$X$ do have
also $Y$'' becomes vacuously true. This convention
is irrespective of whether $Y\neq\emptyset$.

Throughout the paper, ``{\em implications}'' are 
association rules of confidence~1, 
whereas ``{\em partial rules}'' are 
those having a confidence below~1.
When the confidence could be~1 or could
be less, we say simply ``{\em rule}''.

\section{Redundancy Notions}

We start our analysis from one of the notions 
of redundancy defined formally in~\cite{AgYu}.
The notion is employed also, generally with no 
formal definition, in several papers on association 
rules, which subsequently formalize and study 
just some particular cases of redundancy
(e.g.~\cite{KryszPAKDD}, \cite{SaquerDeogun}); 
thus, we have chosen to qualify this
redundancy as ``standard''. We propose also
a small variation, seemingly less restrictive; we have 
not found that variant explicitly defined in 
the literature, but it is quite natural. 

\begin{defi}
\hfill
\begin{enumerate}[(1)]
\item 
\cite{AgYu}
$X_0\to Y_0$ has {\em standard redundancy} 
with respect to $X_1\to Y_1$ 
if the confidence and support of $X_0\to Y_0$ are 
larger than or equal to those of $X_1\to Y_1$, in {\em all}
datasets. 
\item 
$X_0\to Y_0$ has {\em plain redundancy}
 with respect to $X_1\to Y_1$ 
if the confidence of $X_0\to Y_0$ is larger than or equal to
the confidence of $X_1\to Y_1$, in {\em all} datasets. 
\end{enumerate}
\label{plainreddefsA}
\end{defi}

\noindent
Generally, 
we will be interested in applying these definitions 
only to rules $X_0\to Y_0$ where $Y_0\not\subseteq X_0$
since, otherwise, $c(X_0\to Y_0)=1$ for all datasets
and the rule is trivially redundant.
We state and prove separately, for later use,
the following new technical claim:

\begin{lem}
Assume that rule $X_0\to Y_0$ is plainly redundant 
with respect to rule $X_1\to Y_1$, and that 
$Y_0\not\subseteq X_0$. Then $X_0Y_0\subseteq X_1Y_1$. 
\label{suppfromconf}
\end{lem}

\proof
Assume $X_0Y_0\not\subseteq X_1Y_1$, to argue
the contrapositive. Then, we can 
consider a dataset consisting of one transaction $X_0$ and, 
say, $m$ transactions $X_1Y_1$. No transaction includes $X_0Y_0$,
therefore $c(X_0\to Y_0)=0$; however, $c(X_1\to Y_1)$ is 
either 1 or $m/(m+1)$, which can be pushed up as much as desired 
by simply increasing~$m$. Then, plain redundancy does not hold,
because it requires $c(X_0\to Y_0)\geq c(X_1\to Y_1)$ to hold for
all datasets whereas, for this particular dataset, the inequality fails.\qed

The first use of this lemma is to show that plain redundancy is
not, actually, weaker than standard redundancy.

\begin{thm}
Consider any two rules $X_0\to Y_0$ 
and $X_1\to Y_1$ where $Y_0\not\subseteq X_0$.
Then $X_0\to Y_0$ has standard redundancy with respect to $X_1\to Y_1$ 
if and only if
$X_0\to Y_0$ has plain redundancy with respect to $X_1\to Y_1$. 
\label{plainredboundssupport}
\end{thm}

\proof
Standard redundancy clearly implies plain redundancy by definition.
Conversely, plain redundancy implies, first, 
$c(X_0\to Y_0)\geq c(X_1\to Y_1)$ 
by definition and, further, 
$X_0Y_0\subseteq X_1Y_1$ by Lemma~\ref{suppfromconf};
this implies 
in turn $s(X_0\to Y_0) = s(X_0Y_0) \geq s(X_1Y_1) = s(X_1\to Y_1)$,
for all datasets, and standard redundancy holds.\qed

The reference \cite{AgYu} 
also provides two more direct definitions of redundancy:

\begin{defi} 
\hfill
\begin{enumerate}[(1)]
\item 
if $X_1\subset X_0$ and $X_0Y_0 = X_1Y_1$,
rule $X_0\to Y_0$ is
{\em simply redundant} with respect to $X_1\to Y_1$.
\item if $X_1\subseteq X_0$ and $X_0Y_0\subset X_1Y_1$,
rule $X_0\to Y_0$ is
{\em strictly redundant} with respect to $X_1\to Y_1$.
\end{enumerate}
\label{plainreddefsB}
\end{defi}

\noindent
Simple redundancy in \cite{AgYu} is explained as
a potential connection between 
rules that come from the same frequent set, 
in our case $X_0Y_0 = X_1Y_1$. The formal definition
is not identical to our rendering: in its original 
statement in \cite{AgYu}, rule $XZ\to Y$ 
is {\em simply redundant} with respect to $X\to YZ$,
provided that $Z\neq\emptyset$. 
The reason is that, in that reference,
rules are always assumed to have disjoint sides,
and then both formalizations are clearly equivalent.
We do not impose disjointness, so that the natural
formalization of their intuitive explanation is
as we have just stated in Definition~\ref{plainreddefsB}.
The following is very easy to see (and is formally proved 
in~\cite{AgYu}).

\begin{fact}
\cite{AgYu} 
Both simple and strict redundancies imply standard redundancy.
\label{simpleandstrict}
\end{fact}

Note that, in principle, there could possibly be 
many other ways of being redundant beyond simple and strict 
redundancies: we show below, however, that, in essence, this 
is not the case.
We can relate these notions also to 
the {\em cover operator} of \cite{KryszPAKDD}:

\begin{defi} 
\label{KryszCover}
\cite{KryszPAKDD}
Rule $X_1\to Y_1$ covers rule $X_0\to Y_0$ 
when $X_1\subseteq X_0$ and $X_0Y_0\subseteq X_1Y_1$.
\end{defi}

Here, again, the original definition, 
according to which rule $X\to Y$ covers rule $XZ\to Y'$ if
$Z\subseteq Y$ and $Y'\subseteq Y$ (plus some disjointness 
and nonemptiness conditions that we omit) is appropriate
for the case of disjoint sides. The formalization we give
is stated also in \cite{KryszPAKDD} as a property that
characterizes covering. Both simple and strict redundancies 
become thus merged into a single definition.
We observe as well that the same notion is also employed, 
without an explicit name, in~\cite{PhanLuongICDM}.

Again, it should be clear that, 
in Definition~\ref{KryszCover},
the covered rule is indeed plainly redundant: 
whatever the dataset, 
changing from $X_0\to Y_0$ to $X_1\to Y_1$ the
confidence stays equal or increases since, 
in the quotient $\frac{s(XY)}{s(X)}$ 
that defines the confidence of a rule $X\to Y$, 
the numerator cannot decrease from $s(X_0Y_0)$ 
to $s(X_1Y_1)$, whereas the denominator cannot
increase from $s(X_1)$ to $s(X_0)$.
Also, the proposals in 
Definition \ref{plainreddefsB}~and~\ref{KryszCover}
are clearly equivalent:

\begin{fact}
Rule $X_1\to Y_1$ covers rule $X_0\to Y_0$ if and only if
rule $X_0\to Y_0$ is either simply redundant or strictly
redundant with respect 
to $X_1\to Y_1$, or they are equivalent by reflexivity.
\end{fact}

It turns out that all these notions are, in fact,
fully equivalent to plain redundancy; indeed, the following
converse statement is a main new contribution of this section:

\begin{thm}
Assume rule $X_0\to Y_0$ is plainly redundant 
with respect to $X_1\to Y_1$, where $Y_0\not\subseteq X_0$.
Then rule $X_1\to Y_1$ covers rule $X_0\to Y_0$.
\label{plainredcover}
\end{thm}

\proof
By Lemma~\ref{suppfromconf}, $X_0Y_0\subseteq X_1Y_1$. 
To see the other inclusion, $X_1\subseteq X_0$,
assume to the contrary that $X_1\not\subseteq X_0$. 
Then we can consider a dataset in which one transaction 
consists of $X_1Y_1$ and, say, $m$ transactions 
consist of $X_0$. Since $X_1\not\subseteq X_0$, 
these $m$ transactions do not
count towards the supports of $X_1$ or $X_1Y_1$, 
so that the confidence of $X_1\to Y_1$ is~1; 
also, $X_0$ is not adding to the support of $X_0Y_0$
since $Y_0\not\subseteq X_0$. As $X_0Y_0\subseteq X_1Y_1$,
exactly one transaction includes $X_0Y_0$, so that
$c(X_0\to Y_0) = 1/m$, which can be made as low 
as desired. This would contradict plain redundancy.
Hence, plain redundancy implies the two inclusions
in the definition of cover.\qed

Combining the statements so far,
we obtain the following characterization:

\begin{cor}
Consider any two rules $X_0\to Y_0$ 
and $X_1\to Y_1$ where $Y_0\not\subseteq X_0$.
The following are equivalent:
\begin{enumerate}[\em(1)]
\item 
$X_1\subseteq X_0$ and $X_0Y_0\subseteq X_1Y_1$
(that is, rule $X_1\to Y_1$ covers rule $X_0\to Y_0$);
\item 
rule $X_0\to Y_0$ is either simply redundant or strictly
redundant with respect 
to rule $X_1\to Y_1$, or they are equivalent by reflexivity;
\item
rule $X_0\to Y_0$ is plainly redundant with respect to rule $X_1\to Y_1$;
\item
rule $X_0\to Y_0$ is standard redundant with respect to rule $X_1\to Y_1$.
\end{enumerate}
\label{plainredchar}
\end{cor}

\noindent
Marginally, we note here an additional strength of the proofs given.
One could consider attempts at weakening the notion of plain redundancy 
by allowing for a ``margin'' or ``slack'', appropriately bounded, but 
whose value is independent of the dataset, upon comparing confidences.
The slack could be additive or multiplicative: conditions such as 
$c_{\D}(X_0\to Y_0)\geq c_{\D}(X_1\to Y_1)-\delta$ or
$c_{\D}(X_0\to Y_0)\geq \delta c_{\D}(X_1\to Y_1)$, 
for all $\D$ and for $\delta$ independent of $\D$, could be considered. 
However, such approaches do not define different redundancy notions: 
they result in formulations actually equivalent to plain redundancy. 
This is due to the fact that the proofs in Lemma~\ref{suppfromconf}
and Theorem~\ref{plainredcover} show that the gap between
the confidences of rules that do not exhibit redundancy can
be made as large as desired within $(0,1)$. Likewise, if we
fix a confidence threshold $\gamma\in(0,1)$ beforehand and 
use it to define redundancy as
$c_{\D}(X_0\to Y_0)\geq \gamma\dto c_{\D}(X_1\to Y_1)\geq\gamma$
for all $\D$, again an equivalent notion is obtained, 
independently of the concrete value of $\gamma$; whereas, 
for $\gamma=1$, this is, instead, a characterization of 
Armstrong derivability.

\subsection{Deduction Schemes for Plain Redundancy}
\label{dedplain}

{}From the characterization just given, 
we extract now a sound and complete deductive calculus.
It consists of three inference schemes: 
right-hand Reduction ($rR$), where the consequent is diminished;
right-hand Augmentation ($rA$), where the consequent is enlarged; and
left-hand Augmentation ($\ell A$), where the antecedent is enlarged.
As customary in logic calculi, our rendering of each rule means that,
if the facts above the line are already derived, we can immediately
derive the fact below the line.

\smallskip

$(rR)\quad \frac{X\implies Y,\qquad\strut Z\subseteq Y}{X\implies Z}$

\smallskip

$(rA) \quad \frac{X\implies Y}{X\implies XY}$

\smallskip

$(\ell A)\quad \frac{X\implies YZ}{XY\implies Z}$

\smallskip

\noindent
We also allow always to state trivial rules:

\smallskip

$(r\emptyset)\quad \frac{\quad}{X\implies\emptyset}$

\smallskip

\noindent
Clearly, scheme $(\ell A)$ could be stated 
equivalently with $XY\to YZ$ below the line, by $(rA)$:

\smallskip

$(\ell A')\quad \frac{X\implies YZ}{XY\implies YZ}$

\smallskip

In fact, $(\ell A)$ is exactly the simple redundancy from 
Definition~\ref{plainreddefsB} and, in the cases where 
$Y\subseteq X$, it provides a way of dealing with 
one direction of equivalence by reflexivity; 
the other direction is a simple combination of the other two schemes.
The Reduction Scheme $(rR)$ allows us to ``lose''
information from the right-hand side;
it corresponds
to strict redundancy.

As further alternative options, it is easy to see that
we could also join $(rR)$ and $(rA)$ into a single scheme:

\smallskip

$(rA')\quad \frac{X\implies Y,\qquad\strut Z\subseteq XY}{X\implies Z}$

\smallskip

\noindent
but we consider that this option does not
really simplify, rather obscures a bit, 
the proof of our
Corollary~\ref{calculus0complete} below.
Also, we could allow as trivial 
rules $X\to Y$ whenever $Y\subseteq X$, which 
includes the case of $Y=\emptyset$; 
such rules also follow from the calculus given 
by combining $(r\emptyset)$
with $(rA)$ and $(rR)$. 

The following can be derived now 
from Corollary~\ref{plainredchar}:

\begin{cor}
The calculus given is sound and complete for
plain redundancy; that is,
rule $X_0\to Y_0$ is plainly
redundant with respect to rule 
$X_1\to Y_1$ if and only if 
$X_0\to Y_0$ can be derived 
from $X_1\to Y_1$ 
using
the inference schemes $(rR)$, $(rA)$, and $(\ell A)$.
\label{calculus0complete}
\end{cor}

\proof 
Soundness, that is,
all rules derived are plainly redundant, 
is simple to argue by checking that,
in each of the inference schemes,
the confidence of the rule
below the line is greater than or
equal to the confidence of the rule
above the line: these facts are actually the known
statements that each of equivalence by reflexivity,
simple redundancy, and strict redundancy imply plain
redundancy. Also, trivial rules with empty right-hand 
side always hold.
To show completeness,
assume that rule $X_0\to Y_0$ 
is plainly redundant with respect to rule 
$X_1\to Y_1$. If $Y_0\subseteq X_0$,
apply $(r\emptyset)$ and use $(rA)$ 
to copy $X_0$ and, if necessary, $(rR)$ 
to leave just $Y_0$ in the right-hand side. 
If $Y_0\not\subseteq X_0$,
by Corollary~\ref{plainredchar},
we know that this implies that 
$X_1\subseteq X_0$ and $X_0Y_0\subseteq X_1Y_1$.
Now, to infer $X_0\to Y_0$ from $X_1\to Y_1$,
we chain~up applications of our schemes as follows:
$$
    X_1\to Y_1 \, \yields_{(rA)} \,
    X_1\to X_1Y_1 \, \yields_{(rR)} \,
    X_1\to X_0Y_0 \, \yields_{(\ell A)} \,
    X_0\to Y_0
$$
where the second step makes use of the 
inclusion $X_0Y_0 \subseteq X_1Y_1$,
and the last step makes use of the 
inclusion $X_1\subseteq X_0$.
Here, the standard derivation 
symbol $\yields$ denotes derivability 
by application of the scheme indicated
as a subscript.\qed

We note here that \cite{PhanLuongICDM} proposes
a simpler calculus that consists, essentially,
of $(\ell A)$ (called there ``weak left augmentation'')
and $(rR)$ (called there ``decomposition'').
The point is that these two schemes are sufficient
to prove completeness of the ``representative basis''
as given in that reference,
due to the fact that, in that version, the rules 
of the representative basis include
the left-hand side as part of the right-hand side;
but such a calculus is incomplete with respect
to plain redundancy because it 
offers no rule to move items from left to right.

\subsection{Optimum-Size Basis for Plain Redundancy}
\label{subsecAgYubasis}

A basis is a way of providing a shorter list of
rules for a given dataset, with no loss
of information, in the following sense:

\begin{defi}
Given a set of rules $\R$, $\B\subseteq\R$ is 
a {\em complete basis} if every rule of $\R$
is plainly redundant with respect to some rule of $\B$.
\end{defi}

Bases are analogous to covers in functional
dependencies, and we aim at constructing bases
with properties that correspond to minimum size
and canonical covers. The solutions for functional
dependencies, however, are not valid for partial
rules due to the failure of the Armstrong schemes.

In all practical applications, $\R$ is the set of 
all the rules ``mined from'' a given dataset~$\D$ 
at a confidence threshold~$\gamma\in(0,1]$. 
That is, the basis is a set of rules that hold with
confidence at least~$\gamma$ in~$\D$, and such that each
rule holds with confidence at least~$\gamma$ in~$\D$ if
and only if it is plainly redundant with respect to some 
rule of~$\B$; equivalently, the 
rules in~$\R$ 
can be inferred from~$\B$ through the corresponding 
deductive calculus. All along this paper, such a confidence
threshold is denoted $\gamma$, and always~$\gamma>0$.
We will employ two simple but useful definitions.

\begin{defi}
Fix a dataset $\D$.
Given itemsets $Y$ and $X\subseteq Y$, 
$X$ is a $\gamma$-{\em antecedent} for $Y$
if $c(X\to Y) \geq \gamma$, that is, $s(Y)\geq \gamma s(X)$.
\end{defi}

Note that we allow $X=Y$, that is, the set itself as its
own $\gamma$-antecedent; this is just to simplify the
statement of the following rather immediate lemma:

\begin{lem}
If $X$ is a $\gamma$-{\em antecedent} for $Y$ and
$X\subseteq Z\subseteq Y$, then
$X$ is a $\gamma$-{\em antecedent} for $Z$ and
$Z$ is a $\gamma$-{\em antecedent} for $Y$.
\label{chain}
\end{lem}

\proof 
From $X\subseteq Z\subseteq Y$ we have $s(X)\geq s(Z)\geq s(Y)$,
so that $s(Z)\geq s(Y)\geq \gamma s(X) \geq \gamma s(Z)$.
The lemma follows.\qed

We make up for proper antecedents as part of the next notion:

\begin{defi}
Fix a dataset $\D$.
Given itemsets $Y$ and $X\subset Y$ (\emph{proper subset}), 
$X$ is a {\em valid} 
$\gamma$-{\em antecedent} for $Y$ if the following holds:
\begin{enumerate}[(1)]
\item
$X$ is a $\gamma$-antecedent of $Y$,
\item
no proper subset of $X$ is a $\gamma$-antecedent of $Y$, and
\item
no proper superset of $Y$ has $X$ as a $\gamma$-antecedent.
\end{enumerate}
\label{validants}
\end{defi}

\noindent
The basis we will focus on now is constructed from 
each $Y$ and each valid antecedent of $Y$; we consider that this is
the most clear way to define and study it, and we explain 
below why it is essentially identical to two existing, 
independent proposals.

\begin{defi}
Fix a dataset $\D$ and a confidence threshold $\gamma$.
The {\em representative rules} for $\D$ at confidence $\gamma$
are all the rules $X\to Y-X$ for all itemsets $Y$ and for all 
valid $\gamma$-antecedents $X$ of $Y$.
\label{Bcero}
\end{defi}

In the following, we will say 
``let $X\to Y-X$ be a representative rule'' 
to mean 
``let $Y$ be a set having valid $\gamma$-antecedents,
and let $X$ be one of them''; the parameter $\gamma>0$ will
always be clear from the context.
Note that some sets $Y$ may not have valid antecedents, 
and then they do not generate any representative
rules. 

By the conditions on valid antecedents in representative
rules, the following relatively simple but crucial 
property holds; beyond the use 
of our Corollary~\ref{plainredchar}, the argument
follows closely that of related facts in \cite{Krysz}:

\begin{prop}
Let rule $X\to Y-X$ be among the representative rules 
for $\D$ at confidence $\gamma$. Assume that it is
plainly redundant with respect to rule $X'\to Y'$, 
also of confidence at least $\gamma$;
then, they are equivalent by reflexivity and, in case
$X'\cap Y'=\emptyset$, they are the same rule.
\label{equivrepr}
\end{prop}

\proof 
Let $X\to Y-X$ be a representative rule,
so that $X\subset Y$ and $X$ is a valid
$\gamma$-antecedent of $Y$.
By Corollary~\ref{plainredchar},
$X'\to Y'$ must cover $X\to Y-X$:
$X'\subseteq X \subseteq X(Y-X) = Y \subseteq X'Y'$. 
As $c(X'\to Y')\geq\gamma$, $X'$ is
a $\gamma$-antecedent of $X'Y'$. 
We first show that $Y = X'Y'$; assume
$Y \subset X'Y'$, and 
apply Lemma~\ref{chain}
to $X'\subseteq X \subseteq Y \subset X'Y'$: 
$X'$ is also a $\gamma$-antecedent of $Y$,
and the minimality of valid $\gamma$-antecedent $X$ 
gives us $X=X'$.
$X$ is, thus, a $\gamma$-antecedent of $X'Y'$
which properly includes $Y$, contradicting
the third property of valid antecedents.

Hence, $Y = X'Y'$, so that
$X'$ is a $\gamma$-antecedent of $X'Y'=Y$;
but again $X$ is a minimal $\gamma$-antecedent of 
$X'Y'=Y$, so that necessarily $X=X'$, which,
together with $X'Y'=Y=XY$, proves equivalence 
by reflexivity. 
Under the additional condition 
$X'\cap Y'=\emptyset$, 
both rules coincide as per Proposition~\ref{disjointness}.\qed

It easily follows that our definition is equivalent
to the definition given in \cite{KryszPAKDD}, except
for a support bound that we will explain later;
indeed, we will show in Section~\ref{suppbound} that
all our results carry over when a support bound
is additionally enforced.

\begin{cor}
Fix a dataset $\D$ and a confidence threshold $\gamma$.
Let $X\subset Y$. The following are equivalent:
\begin{enumerate}[\em(1)]
\item
Rule $X\to Y-X$ is among the representative rules 
for $\D$ at confidence~$\gamma$;
\item
\cite{KryszPAKDD}
$c(X\to Y-X)\geq\gamma$ and
there does not exist any other rule $X'\to Y'$ 
with $X'\cap Y'=\emptyset$, 
of confidence at least $\gamma$ in $\D$,
that covers $X\to Y-X$.
\end{enumerate}
\label{charreprbasisK}
\end{cor} 

\proof
Let rule $X\to Y-X$ be among the representative rules 
for $\D$ at confidence~$\gamma$, and let rule $X'\to Y'$ 
cover it, while being also of confidence at least $\gamma$
and with $X'\cap Y'=\emptyset$. Then, by
Corollary~\ref{plainredchar} $X'\to Y'$ makes 
$X\to Y-X$ plainly redundant, and by 
Proposition~\ref{equivrepr} they must coincide.
To show the converse, we must see that $X\to Y-X$
is a representative rule under the conditions given.
The fact that $c(X\to Y-X)\geq\gamma$ gives that 
$X$ is a $\gamma$-antecedent of $Y$, and we must see its
validity. Assume that a proper subset $X'\subset X$ is 
also a $\gamma$-antecedent of $Y$: then the rule 
$X'\to Y-X'$ would be a different rule of confidence
at least~$\gamma$ covering 
$X\to Y-X$, which cannot be. Similarly, assume that 
$X$ is a $\gamma$-antecedent of $Y'$ where $Y\subset Y'$:
then the rule $X\to Y'-X$ would be a different rule 
of confidence at least~$\gamma$ covering 
$X\to Y-X$, which cannot be either.\qed

Similarly, and with the same proviso regarding support,
our definition is equivalent to the ``essential rules'' 
of \cite{AgYu}.
There, the set of minimal $\gamma$-antecedents
of a given itemset is termed its ``boundary''.
The following statement is also easy to prove:

\begin{cor}
Fix a dataset $\D$ and a confidence threshold $\gamma$.
Let $X\subseteq Y$. The following are equivalent:
\begin{enumerate}[\em(1)]
\item
Rule $X\to Y-X$ is among the representative rules 
for $\D$ at confidence~$\gamma$;
\item
\cite{AgYu}
$X$ is in the boundary of $Y$ but is not in the
boundary of any proper superset of $Y$; that is,
$X$ is a minimal 
$\gamma$-antecedent of $Y$ but is not a minimal
$\gamma$-antecedent of any itemset strictly
containing $Y$.
\end{enumerate}
\label{charreprbasisAY}
\end{cor} 

\proof
If $X\to Y-X$ is among the representative rules, 
$X$ must be a minimal $\gamma$-antecedent of $Y$
by the conditions of valid antecedents; also,
$X$ is not a $\gamma$-antecedent at all (and, thus, 
not a minimal $\gamma$-antecedent)
of any $Y'$ properly including $Y$. 
Conversely, assume that
$X$ is in the boundary of $Y$ but is not in the
boundary of any proper superset of $Y$; first,
$X$ must be a minimal $\gamma$-antecedent of $Y$
so that the first two conditions of valid $\gamma$-antecedents
hold. Assume that $X\to Y-X$ is not among the 
representative rules; the third property must fail,
and $X$ must be a $\gamma$-antecedent of some $Y'$
with $Y\subset Y'$. Our hypotheses tell us that
$X$ is not a \emph{minimal} $\gamma$-antecedent of $Y'$.
That is, there is a proper subset $X'\subset X$
that is also a $\gamma$-antecedent of $Y'$.
It suffices to apply Lemma~\ref{chain} to
$X'\subset X \subseteq Y\subset Y'$ to reach
a contradiction, since it implies that
$X'$ is a $\gamma$-antecedent of $Y$ and therefore
$X$ would not be a minimal $\gamma$-antecedent of~$Y$.\qed

The representative rules are indeed a basis:

\begin{fact}
(\cite{AgYu}, \cite{KryszPAKDD})
Fix a dataset $\D$ and a confidence threshold $\gamma$, 
and consider the set of representative rules 
constructed from $\D$; it is a complete basis: 
\begin{enumerate}[(1)]
\item 
all the representative rules hold with 
confidence at least $\gamma$;
\item 
all the rules of confidence at least $\gamma$
in $\D$ are plainly redundant with respect to 
the representative rules.
\end{enumerate}
\label{B0iscomplete}
\end{fact}

\noindent
The first part follows directly from the use of
$\gamma$-antecedents as left-hand sides of 
representative rules. For the second part,
also almost immediate,
suppose $c(X\to Y)\geq\gamma$, and let $Z=XY$;
since $X$ is now a $\gamma$-antecedent of $Z$,
it must contain a minimal $\gamma$-antecedent of $Z$,
say $X'\subseteq X$. Let $Z'$ be the largest superset 
of $Z$ such that $X'$ is still a $\gamma$-antecedent 
of $Z'$. Thus, $X'\to Z'-X'$ is among the representative 
rules and covers $X\to Y$.
Small examples of the construction of representative rules
can be found in the same references; we also provide one below. 

An analogous fact is proved in \cite{PhanLuongICDM} 
through an incomplete deductive calculus
consisting of the schemes that we have called
$(lA)$ and $(rR)$,
and states that every rule of confidence at least~$\gamma$
can be inferred from the representative rules by
application of these two inference schemes. Since
representative rules 
in the formulation of \cite{PhanLuongICDM}
have a right-hand side
that includes the left-hand side, this inference
process does not need to employ $(rA)$.

Now we can state and prove the 
most interesting novel property 
of this basis, which again follows 
from our main result in this section, 
Corollary~\ref{plainredchar}. 
As indicated, representative rules
were known to be irredundant with 
respect to simple and strict redundancy
or, equivalently, with respect to covering. But, 
for standard redundancy, in principle 
there was actually the possibility 
that some other basis, constructed
in an altogether different form, 
could have less rules. We can state 
and prove now that this is not so:
there is absolutely no other way 
of constructing a basis smaller than 
this one, while preserving completeness
with respect to plain redundancy,
because it has absolutely minimum size 
among all complete bases. Therefore, 
in order to find smaller bases,
a notion of redundancy more
powerful than plain (or standard) 
redundancy is unavoidably necessary.

\begin{thm}
Fix a dataset $\D$, and let $\R$ be the set of rules
that hold with confidence~$\gamma$ in $\D$.
Let $\B'\subseteq\R$ be an arbitrary basis, complete so that
all the rules in $\R$ are plainly redundant with respect to $\B'$.
Then, $\B'$ must have at least as many rules as 
the representative rules. Moreover, if the rules
in $\B'$ are such that antecedents and consequents
are disjoint, then all the representative rules 
belong to $\B'$.
\label{zerominimality}
\end{thm}

\proof
By the assumed completeness of $\B'$, 
each representative rule $X\to Y-X$ 
must be redundant with respect to 
some rule $X'\to Y'\in\B'\subseteq\R$.
By Corollary~\ref{plainredchar}, 
$X'\to Y'$ covers $X\to Y-X$.
Then Proposition~\ref{equivrepr} applies:
they are equivalent by reflexivity.
This means $X = X'$ and $Y = X'Y'$,
hence $X'\to Y'$ uniquely identifies
which representative rule it covers,
if any; hence, $\B'$ needs, at~least, as
many rules as the number of representative
rules. Moreover, 
as stated also in Proposition~\ref{equivrepr},
if the disjointness condition
$X'\cap Y' = \emptyset$ holds,
then both rules coincide.\qed

\begin{exa}
We consider a small example consisting of 12 transactions, where 
there are actually only 7 itemsets, but some of them are repeated
across several transactions. We can simplify our study as follows:
if $X$ is not a closed set for the dataset, that is, if it has
some superset $X'\supset X$ with the same support, then clearly
it has no valid $\gamma$-antecedents 
(see also Fact~\ref{Bcerofromclosures} below);
thus we concentrate on closed sets.
Figure~\ref{smallex} shows the example dataset and the corresponding 
\hbox{(semi-)lattice} of closures, depicted as a Hasse diagram (that~is, 
transitive edges have been removed to clarify the drawing); edges stand 
for the inclusion relationship.

\begin{figure}[t]
\begin{center}
\includegraphics[width=14cm]{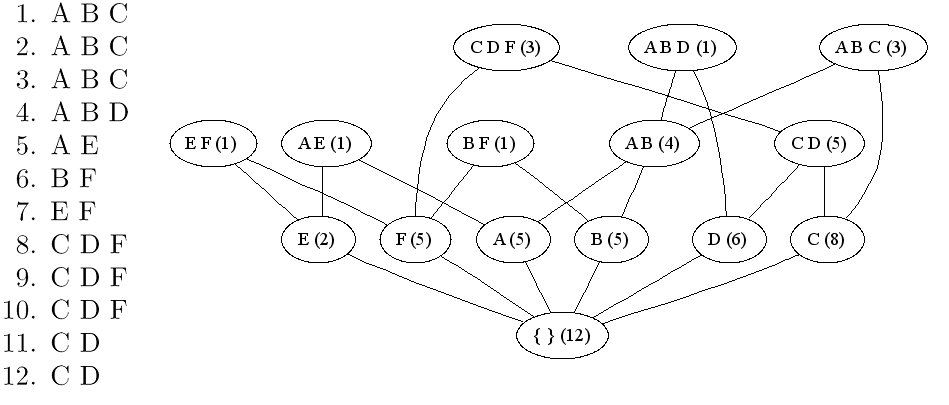}
\end{center}
\caption{Closed itemsets for a small example}
\label{smallex}
\end{figure}

For this example, the implications can be summarized by six rules, 
namely, $AC\dto B$, 
$BC\dto A$, 
$AD\dto B$, 
$BD\dto A$, 
$CF\dto D$, 
and $DF\dto C$,
which are also the
representative rules at confidence~1. 
At confidence $\gamma=0.75$, 
we find that, first, the left-hand sides of the
six implications are still valid $\gamma$-antecedents
even at this lower confidence, so that the implications
still belong to the representative basis. 
Then, we see that two of the closures, $ABC$ and $CD$, 
have additionally one valid
$\gamma$-antecedent each, whereas $AB$ has two.
The following four rules hold: $A\to B$, $B\to A$, 
$AB\to C$, and $D\to C$. These four rules,
jointly with the six implications indicated, 
constitute exactly 
the ten representative rules at confidence~0.75. 
\end{exa}

\section{Closure-Based Redundancy}

Theorem~\ref{zerominimality} in the previous section 
tells us that, 
for plain redundancy, the absolute limit of
a basis at any given confidence threshold
is reached by the 
set of representative rules.
Several studies, prominently \cite{Zaki}, 
have put forward a different notion of redundancy;
namely, they give a separate role to the
full-confidence implications, often through
their associated closure operator. Along this way,
one gets a stronger notion of redundancy and, 
therefore, a possibility that smaller bases can be constructed.

Indeed, implications can be summarized better, because they allow for
Transitivity and Augmentation to apply in order to find redundancies;
moreover, they can be combined in certain forms of transitivity with
partial rules: as a simple example, if $c(X\to Y)\geq\gamma$ 
and $c(Y\to Z)=1$, that is, if a fraction $\gamma$ or more of the support
of $X$ has $Y$ and {\em all} the transactions containing $Y$ do have
$Z$ as well, clearly this implies that $c(X\to Z)\geq\gamma$.
Observe, however, that the directionality is relevant:
from $c(X\to Y)=1$ and $c(Y\to Z)\geq\gamma$ we infer 
nothing about $c(X\to Z)$, since the high confidence 
of $Y\to Z$ might be due to a large number of transactions
that do not include $X$.

We will need some notation about closures. Given 
a dataset $\D$, the closure operator associated 
to $\D$ maps each itemset $X$ to the largest 
itemset $\cls{X}$ that contains $X$ and has 
the same support as $X$ in $\D$: 
$s(X) = s(\cls{X})$, and $\cls{X}$ is as
large as possible under this condition. 
It is known and easy to prove that $\cls{X}$
exists and is unique. 
Implications that hold in the dataset correspond
to the closure operator 
(\cite{GW}, 
\cite{GD}, 
\cite{PasBas}, 
\cite{Wild},
\cite{Zaki}): 
$c(X\to\cls{X}) = 1$, and $\cls{X}$ is as
large as possible under this condition. 
Equivalently, the closure
of itemset $X$ is the intersection of all the
transactions that contain $X$; this is
because $X\subseteq\cls{X}$ implies that all
transactions counted for the support of $\cls{X}$
are counted as well for the support of $X$, hence,
if the support counts coincide they must count
exactly the same transactions. 

Along this section, as in \cite{PasBas}, we denote 
full-confidence implications using
the standard logic notation $X_0\dto Y_0$; thus,
$X_0\dto Y_0$ if and only if $Y_0\subseteq\cls{X_0}$.

A basic fact from the theory of Closure Spaces is that
closure operators are characterized by three properties:
extensivity ($X\subseteq\cls{X}$), idempotency
($\cls{\cls{X}} = \cls{X}$), and monotonicity
(if $X\subseteq Y$ then $\cls{X}\subseteq\cls{Y}$).
As an example of the use of these properties, we note
the following simple consequence for later use:

\begin{lem}
$XY\subseteq\cls{X}Y\subseteq\cls{X}\,\cls{Y}\subseteq\cls{XY}$,
and 
$\cls{XY} = \cls{\cls{X}Y} = \cls{\cls{X}\,\cls{Y}} = \cls{\cls{XY}} = \cls{XY}$.
\label{trivial}
\end{lem}

We omit the immediate proof.
A set is closed if it coincides with its closure. 
Usually we speak of the {\em lattice} of closed sets
(technically it is just a semilattice but it allows
for a standard transformation into a lattice \cite{DaPr}).
When $\cls{X} = Y$ we also say that $X$ is a 
generator of $Y$; if the closures of all proper
subsets of $X$ are different from $Y$, we say
that $X$ is a minimal generator. Note that some
references use the term ``generator'' to mean our
``minimal generator''; we prefer to make explicit
the minimality condition in the name. In some
works, often database-inspired, minimal generators 
are termed sometimes ``keys''.
In other works, often matroid-inspired, 
they are termed also ``free sets''.
Our definition says explicitly that
$s(X) = s(\cls{X})$. We will make liberal use of this fact, 
which is easy to check also with other 
existing alternative definitions of the 
closure operator, as stated in \cite{PasBas}, \cite{Zaki}, 
and others. Several quite good algorithms exist to find the 
closed sets and their supports (see section~4 of~\cite{CegRod}). 

Redundancy based on closures is a natural
generalization of equivalence by reflexivity;
it works as follows
(\cite{Zaki}, see also \cite{Krysz} and section~4 in~\cite{PasBas}): 

\begin{lem}
Given a dataset and the corresponding closure operator, 
two partial rules $X_0\to Y_0$ and $X_1\to Y_1$ 
such that $\cls{X_0} = \cls{X_1}$ and
$\cls{X_0Y_0} = \cls{X_1Y_1}$ have the same
support and the same confidence.
\label{closredCCI}
\end{lem}

The rather immediate reason is that 
$s(X_0) = s(\cls{X_0}) = s(\cls{X_1}) = 
s(X_1)$, and $s(X_0Y_0) = s(\cls{X_0Y_0}) = 
s(\cls{X_1Y_1}) = s(X_1Y_1)$.
Therefore, groups of rules sharing the
same closure of the antecedent, and the
same closure of the union of antecedent
and consequent, give cases of redundancy.
On account of these properties, there are 
some proposals of basis constructions
from closed sets in the literature, 
reviewed below. But the first fact that
we must mention to relate the closure
operator with our explanations so far
is the following:

\begin{fact}
\cite{KryszRSCTC}
Let $X\to Y-X$ be a representative rule as per Definition~\ref{Bcero}. 
Then $Y$ is a closed set and $X$ is a minimal generator. 
\label{Bcerofromclosures}
\end{fact}

The proof is direct from Definitions \ref{validants}~and~\ref{Bcero},
and can be found in 
\cite{KryszRSCTC}, 
\cite{Krysz}, 
\cite{PhanLuongICDM}.
These references employ this property
to improve on the earlier algorithms 
to compute the representative rules, 
which considered all the frequent sets, 
by restricting the exploration to closures
and minimal generators. Also the authors of
\cite{SaquerDeogun} 
do the same, seemingly unaware that the algorithm
in \cite{KryszRSCTC} already works just with
closed itemsets.
Fact~\ref{Bcerofromclosures} 
may shed doubts on whether closure-based 
redundancy actually can lead to smaller bases. 
We prove that this is sometimes the case,
due to the fact that the redundancy notion
itself changes, and allows for a form of
Transitivity, which we show can take again the 
form of a deductive calculus. Then, we will
be able to refine the notion of valid antecedent 
of the previous section and provide a basis
for which we can prove that it has
the smallest possible size
among the bases for partial rules,
with respect to closure-based completeness.
That is, we will reach the limit 
of closure-based redundancy in the same
manner as we did for standard redundancy
in the previous section.

\subsection{Characterizing Closure-Based Redundancy}\label{subscharclosred}

Let $\B$ be the set of implications
in the dataset $\D$; alternatively, $\B$ can be any
of the bases already known for implications in a dataset.
In our empirical validations below
we have used as $\B$ the
Guigues-Duquenne basis, or GD-basis, that has been proved to be
of minimum size \cite{GD}, \cite{Wild}.
An apparently popular and interesting alternative,
that has been rediscovered over and over in 
different guises, is the so-called
{\em iteration-free basis} of \cite{Wild}, which 
coincides with the proposal in \cite{PT} and 
with the {\em exact min-max basis} of \cite{PasBas}
(also called sometimes {\em generic basis} \cite{Krysz}); 
because of Fact~\ref{Bcerofromclosures},
it coincides exactly also with
the representative rules of confidence~1, that is: 
implications that
are not plainly redundant with any other implication
according to Definition~\ref{plainreddefsA}.
Also, it coincides with the ``closed-key basis''
for frequent sets in \cite{PhanLuongDawak}, which
in principle is not intended as a basis for rules,
and has a different syntactic sugar, but differs
in essence from the iteration-free basis only
in the fact that the support of each rule is
explicitly recorded together with it. 

Closure-based redundancy takes into account $\B$
as follows:

\begin{defi}
Let $\B$ be a set of implications.
Partial rule $X_0\implies Y_0$ has {\em closure-based
redundancy relative to $\B$} with respect to rule 
$X_1\implies Y_1$, denoted
$\B,\{X_1\implies Y_1\}\models X_0\implies Y_0$,
if any dataset $\D$ in which all the rules in $\B$ 
hold with confidence 1 gives
$c_{\D}(X_0\to Y_0) \geq c_{\D}(X_1\to Y_1)$.
\label{closredundancy}
\end{defi}

In some cases, it might happen that the dataset at hand
does not satisfy any nontrivial rule with confidence 1;
then, this notion will not be able to go beyond plain redundancy.
However, it is usual that some full-confidence rules do hold, 
and, in these cases, as we shall see, closure-based 
redundancy may give more economical bases.
More generally, all our results only depend on the implications
reaching indeed full confidence in the dataset; but they are not 
required to capture all of these:
the implications in $\B$ (with their consequences according to
the Armstrong schemes) could constitute just a part of the
full-confidence rules in the dataset. In particular, plain
redundancy reappears by choosing $\B=\emptyset$, whether the
dataset satisfies or not any full-confidence implication. 

We continue our study by showing a necessary and sufficient 
condition for closure-based redundancy, along the same
lines as the one in the previous section.

\begin{thm}
Let $\B$ be a set of exact rules, with associated closure operator
mapping each itemset $Z$ to its closure $\cls{Z}$. Let $X_0\implies Y_0$ be
a rule not implied by~$\B$, that is, where $Y_0\not\subseteq\cls{X_0}$.
Then, the following are equivalent:
\begin{enumerate}[\em(1)]
\item 
$X_1\subseteq\cls{X_0}$ and $X_0Y_0\subseteq\cls{X_1Y_1}$;
\item 
$\B,\{X_1\implies Y_1\}\models X_0\implies Y_0$.
\end{enumerate}
\label{closredchar}
\end{thm}

\proof
The direct proof is simple: the inclusions given imply that
$s(X_1)\geq s(\cls{X_0}) = s(X_0)$ and 
$s(X_0Y_0)\geq s(\cls{X_1Y_1}) = s(X_1Y_1)$;
then $c(X_0\to Y_0) = \frac{s(X_0Y_0)}{s(X_0)} 
\geq \frac{s(X_1Y_1)}{s(X_1)} = c(X_1\to Y_1)$.

Conversely, for $Y_0\not\subseteq\cls{X_0}$, 
we argue that, if either of 
$X_1\subseteq\cls{X_0}$ and 
$X_0Y_0\subseteq\cls{X_1Y_1}$ fails,
then there is a dataset where $\B$ holds 
with confidence 1 and $X_1\to Y_1$ holds with high
confidence but the confidence of $X_0\to Y_0$ is low.

We observe first that, in order to satisfy $\B$, it suffices
to make sure that all the transactions in the dataset we
are to construct are closed sets according to the closure
operator corresponding to $\B$.

Assume now that $X_1\not\subseteq\cls{X_0}$: then a dataset
consisting only of one or more transactions~with itemset
$\cls{X_0}$ satisfies (vacuously) $X_1\to Y_1$ with
confidence 1 but, given that 
$Y_0\not\subseteq\cls{X_0}$, leads to confidence zero 
for $X_0\to Y_0$. It is also possible to argue 
without resorting to vacuous satisfaction: simply take one
transaction consisting of $\cls{X_1Y_1}$ and, in case this
transaction satisfies $X_0\to Y_0$, obtain as low a confidence
as desired for $X_0\to Y_0$ by adding as many transactions
$\cls{X_0}$ as necessary; these will not change the
confidence of $X_1\to Y_1$ since $X_1\not\subseteq\cls{X_0}$.

Then consider the case where $X_1\subseteq\cls{X_0}$, whence
the other inclusion fails: $X_0Y_0\not\subseteq\cls{X_1Y_1}$.
Consider a dataset of, say, $n$ transactions, where
one transaction consists of the itemset $\cls{X_0}$
and $n-1$ transactions consist of the itemset $\cls{X_1Y_1}$.
The confidence of $X_1\to Y_1$ is at least $\frac{n-1}{n}$,
which can be made as close to 1 as desired by 
increasing $n$, whereas the presence of at least one $X_0$ and no 
transaction at all containing $X_0Y_0$ gives confidence zero
to $X_0\to Y_0$. Thus, in either case, we see that 
redundancy does not hold.\qed

\subsection{Deduction Schemes for Closure-Based Redundancy}
\label{redundcalculus}

We provide now a stronger calculus that is sound and complete 
for this more general case of closure-based redundancy. 
For clarity, we chose
to avoid the closure operator in our deduction schemes, writing
instead explicitly each implication. 

Our calculus for closure-based redundancy consists of four inference 
schemes, each of which reaches a partial rule from premises including 
a partial rule. Two of the schemes correspond to variants of Augmentation, 
one for enlarging the antecedent, the other for enlarging the consequent. 
The other two correspond to composition with an implication, one in the 
antecedent and one in the consequent: a form of controlled transitivity. 
Their names $(rA)$, $(\ell A)$, $(rI)$, and $(\ell I)$ indicate whether 
they operate at the right or left-hand side and whether their effect is 
Augmentation or composition with an Implication.

\smallskip

$(rA) \quad \frac{X\to Y,\strut\qquad X\dto Z}{X\to YZ}$

\smallskip

$(rI)\quad \frac{X\to Y,\strut\qquad Y\dto Z}{X\to Z}$

\smallskip

$(\ell A)\quad \frac{X\to YZ}{XY\to Z}$

\smallskip

$(\ell I)\quad \frac{X\to Y,\strut\qquad Z\subseteq X,\qquad Z\dto X}{Z\to Y}$

\smallskip

\noindent
Again we allow to state rules with empty right-hand side directly: 

\smallskip

$(r\emptyset)\quad \frac{\quad}{X\implies\emptyset}$

\smallskip

\noindent
Alternatively, we could state trivial rules with a subset
of the left-hand side at the right-hand side. Note that this opens 
the door to using 
$(rA)$ with an empty $Y$, and this allows us to ``downgrade'' an implication 
into the corresponding partial rule. Again, $(\ell A)$ could be stated 
equivalently as $(\ell A')$ like in Section~\ref{dedplain}.
In fact, the whole connection with the simpler calculus 
in Section~\ref{dedplain} should be easy to 
understand: first, observe that the $(\ell A)$ rules are identical.
Now, if implications are not considered separately, the closure
operator trivializes to identity, $\cls{Z}=Z$ for every $Z$, and
the only cases where we know that $X_1\dto Y_1$ are those where 
$Y_1\subseteq X_1$; we see that $(rI)$ corresponds, 
in that case, to $(rR)$, whereas the $(rA)$ schemes only differ on 
cases of equivalence by reflexivity. Finally, in that case 
$(\ell I)$ becomes fully trivial since $Z\dto X$ becomes 
$X\subseteq Z$ and, together with $Z\subseteq X$, leads to $X=Z$: 
then, the partial rules above and below the line would coincide.

Similarly to the plain case, there exists an alternative 
deduction system, more compact, whose equivalence
with our four schemes is rather easy to see. 
It consists of just two forms of combining 
a partial rule with an implication:

\smallskip

$(rI') \quad \frac{X\to Y,\strut\qquad XY\dto Z}{X\to Z}$

\smallskip

$(\ell I')\quad \frac{X\to Y,\strut\qquad Z\subseteq XY,\qquad Z\dto X}{Z\to Y}$

\smallskip

However, in our opinion, the use of these schemes in our
further developments is less intuitive, so we keep working 
with the four schemes above.

In the remainder of this section, we denote as
$\B,\{X\to Y\}\yields X'\to Y'$ 
the fact that, in the presence of 
the implications in the set $\B$,
rule $X'\to Y'$ can be derived from 
rule $X\to Y$ using zero or more
applications of the four deduction schemes;
along such a derivation, any rule of $\B$
(or derived from $\B$ by the Armstrong schemes)
can be used whenever an implication of the
form $X\dto Y$ is required.

\subsection{Soundness and Completeness} 
\label{clocalcsoundcompl}

We can characterize the deductive power 
of this calculus as follows: it is sound
and complete with respect to the notion 
of closure-based redundancy; that is,
all the rules it can prove are redundant,
and all the redundant rules can be proved:

\begin{thm}
Let $\B$ consist of implications. Then,
$\B,\{X_1\implies Y_1\}\yields X_0\implies Y_0$
if and only if 
rule $X_0\to Y_0$ has closure-based
redundancy relative to $\B$
with respect to rule $X_1\to Y_1$:
$\B,\{X_1\implies Y_1\}\models X_0\implies Y_0$.
\label{calculussoundcomplete}
\end{thm}

\proof
Soundness corresponds to the fact
that every rule derived is redundant: it suffices
to prove it individually for each scheme; 
the essentials of some of these 
arguments are also found in the literature. 
For $(rA)$, 
 the inclusions $XY\subseteq XYZ\subseteq\cls{XY}$
 prove that the partial rules above and below
 the line have the same confidence. For $(rI)$, 
 one has $XZ\subseteq X\cls{Y}\subseteq\cls{XY}$,
 thus $s(XZ)\geq s(XY)$ and the confidence of 
 the rule below the line is at least that of the
 one above, or possibly
greater. Scheme $(\ell A)$ is 
unchanged from the previous section. 
 Finally, for
 $(\ell I)$, we have $Z\subseteq X\subseteq\cls{Z}$
 so that $s(Z) = s(X)$, and $ZY\subseteq XY$ so that
 $s(ZY)\geq s(XY)$, and again
 the confidence of the rule below the line is
 at least the same as the confidence of the one above.

To prove completeness, we must see that all
redundant rules can be derived. We assume
$\B,\{X_1\implies Y_1\}\models X_0\implies Y_0$
and resort to Theorem~\ref{closredchar}:
we know that the inclusions $X_1\subseteq\cls{X_0}$ 
and $X_0Y_0\subseteq\cls{X_1Y_1}$ must hold.
From Lemma~\ref{trivial}, we have that
$\cls{X_0}Y_0\subseteq\cls{X_1Y_1}$.

Now we can write a derivation in our calculus,
taking into account these inclusions, as follows:
$$
    X_1\to Y_1 \yields_{(rA)} 
    X_1\to X_1Y_1 \yields_{(rI)} 
    X_1\to \cls{X_0}Y_0 \yields_{(\ell A)} 
    \cls{X_0}\to Y_0 \yields_{(\ell I)} 
    X_0\to Y_0
$$
Thus, indeed the redundant rule is derivable,
which proves completeness.\qed

\subsection{Optimum-Size Basis for Closure-Based Redundancy}
\label{mybasissect}

In a similar way as we did for plain redundancy,
we study here bases corresponding to closure-based
redundancy. 

Since the implications become ``factored out'' thanks
to the stronger notion of redundancy, 
we can focus on the partial rules. 
A formal definition 
of completeness for a basis is, therefore, as follows:

\begin{defi}
Given a set of partial rules $\R$ 
and a set of implications $\B$, 
{\em closure-based completeness}
of a set of partial rules 
$\B'\subseteq\R$ holds 
if every partial rule of $\R$
has {\em closure-based
redundancy relative to $\B$} 
with respect to some rule of $\B'$.
\end{defi}

Again $\R$ is intended to be the set of 
all the partial rules ``mined from'' a given dataset~$\D$ 
at a confidence threshold~$\gamma<1$ (recall
that always~$\gamma>0$), whereas $\B$ is
intended to be the subset of rules in $\R$
that hold with confidence~1 in $\D$ 
or, rather, a basis for these implications. 
There exist several proposals for constructing
bases while taking into account the implications
and their closure operator. We use the same
intuitions and {\em modus operandi} to add a new 
proposal which, conceptually, departs only 
slightly from existing ones. Its main merit
is not the conceptual novelty of the basis
itself but the mathematical proof that it
achieves the minimum possible size for a
basis with respect to closure-based
redundancy, and is therefore at most as
large as any alternative basis and, 
in many cases, smaller than existing ones.

Our new basis is constructed as follows.
For each closed set $Y$, we will consider 
a number of closed sets $X$ properly included 
in $Y$ as candidates to act as antecedents:

\begin{defi}
Fix a dataset $\D$, and consider the closure operator
corresponding to the implications that hold in $\D$
with confidence~1.
For each closed set $Y$, a closed proper subset $X\subset Y$ 
is a {\em basic $\gamma$-antecedent} 
if the following holds:
\begin{enumerate}[(1)]
\item 
$X$ is a $\gamma$-antecedent of $Y$: $s(Y)\geq \gamma s(X)$;
\item 
no proper closed subset of $X$ is a $\gamma$-antecedent of $Y$, and
\item
no proper closed superset of $Y$ has $X$ as a $\gamma$-antecedent.
\end{enumerate}
\end{defi}

\noindent
Basic antecedents follow essentially the same pattern 
as the valid antecedents (Definition~\ref{validants}),
but restricted to closed sets only, that is, 
instead of minimal antecedents, we pick just 
minimal {\em closed} antecedents.
Then we can use them as before:

\begin{defi}
Fix a dataset $\D$ and a confidence threshold $\gamma$.
\begin{enumerate}[(1)]
\item 
The basis $\myB$ 
consists of all the rules
$X\to Y-X$ for all closed sets $Y$ and all 
basic $\gamma$-antecedents $X$ of $Y$.
\item 
A {\em minmax variant} of the basis
$\myB$ is obtained by replacing each
left-hand side in $\myB$ by a minimal 
generator: that is, for a closed set $Y$, 
each rule $X\to Y-X$ becomes 
$X'\to Y-X$ for one minimal generator $X'$ 
of the (closed) basic $\gamma$-antecedent $X$.
\item 
A {\em minmin variant} of the basis
$\myB$ is obtained by replacing 
by a minimal generator both the
left-hand and the right-hand sides in $\myB$:
for each closed set $Y$ and each basic
$\gamma$-antecedent $X$ of $Y$, 
the rule $X\to Y-X$ becomes 
$X'\to Y'-X$ where $Y'$
is chosen a minimal generator of $Y$ and $X'$
is chosen a minimal generator of $X$.
\end{enumerate}
\label{mybasis}
\end{defi}

\noindent
The variants are defined only for the purpose of
discussing the relationship to previous works along
the next few paragraphs; generally, we will use only 
the first version of $\myB$. Note the following: in a minmax 
variant, at the time of substituting a generator
for the left-hand side closure, in case we consider a rule 
from $\myB$ that has a left-hand side with several minimal 
generators, only one of them is to be used. Also,
all of $X$ (and not only $X'$) can be removed
from the right-hand side: $(rA)$ can be used
to recover it.

The basis $\myB$ is uniquely determined by the dataset and
the confidence threshold, but the variants can be constructed,
in general, in several ways, because each closed set in the rule
may have several minimal generators,
and even several different generators of minimum size. 
We can see the variants as applications of our deduction schemes. 
The result of substituting
a generator for the left-hand side 
of a rule is equivalent to the rule itself:
in one direction it is exactly scheme $(\ell I)$, and in the 
other is a chained application of $(rA)$ to add the closure to
the right-hand side and $(\ell A)$ to put it back in the left-hand 
side. Substituting a generator for the right-hand side corresponds
to scheme $(rI)$ in both directions. 

The use of generators instead of closed sets in
the rules is discussed in several references, 
such as \cite{PasBas} or \cite{Zaki}. In the 
style of \cite{PasBas}, we would consider a 
minmax variant, which allows one to show to 
the user minimal sets of antecedents together 
with all their nontrivial consequents.
In the style of \cite{Zaki}, we would consider 
a minmin variant, thus reducing the total number
of symbols if minimum-size generators are used, since
we can pick any generator. Each of these known bases
incurs a risk of picking more than one minimum generator
for the same closure as left-hand sides of rules with
the same closure of the right-hand side: this is where
they may be (and, in actual cases, have been empirically 
found to be) larger than $\myB$, because, in a sense,
they would keep in the basis all the variants.
Facts analogous to Corollaries 
\ref{charreprbasisK}~and~\ref{charreprbasisAY}
hold as well if the closure condition
is added throughout, and provide further 
alternative definitions of the same basis.
We use one of them in our experimental setting,
described in Section~\ref{experim}.
We now see that this set of rules entails exactly the rules 
that reach the corresponding confidence threshold in the dataset:

\begin{thm} 
Fix a dataset $\D$ and a confidence threshold $\gamma$.
Let $\B$ be any basis for 
implications that hold with confidence~1 in~$\D$.
\begin{enumerate}[\em(1)]
\item All the rules in $\myB$ hold with 
confidence at least $\gamma$.
\item $\myB$ is a complete basis
for the partial rules under
closure-based redundancy. 
\end{enumerate}
\label{newsoundcomplete}
\end{thm}

\proof
All the rules in $\myB$
must hold indeed because all the left-hand sides 
are actually $\gamma$-antecedents. To prove that all the
partial
rules that hold are entailed by rules in~$\myB$, assume 
that indeed $X\to Y$ holds with confidence $\gamma$, 
that is, $s(\cls{XY}) = s(XY)\geq \gamma s(X)$; thus $X$ is a 
$\gamma$-antecedent of $\cls{XY}$. If $Y\subseteq\cls{X}$,
then $c(X\to Y)=1$ and the implication will follow from $\B$;
we have to discuss only the case where $Y\not\subseteq\cls{X}$,
which implies that $\cls{X}\subset\cls{XY}$. Consider the
family of closed sets that include $XY$ and have
$X$ as $\gamma$-antecedent; it is a nonempty family,
since $\cls{XY}$ fulfills these conditions. Pick $Z$
maximal in that family. Then $\cls{X}\subset \cls{Z} = Z$
since $X\subseteq Z$ and $\cls{X}\subset\cls{XY}\subseteq Z$.
Now, $\cls{X}$ is a $\gamma$-antecedent of $Z$,
but not of any strictly larger closed itemset.
Also, any subset of $\cls{X}$ is a proper subset of $Z$. 

Let $X'\subseteq\cls{X}$ be closed, a $\gamma$-antecedent of $Z$, 
and minimal with respect to these properties; assume that $X'$
is a $\gamma$-antecedent of a closed set $Z'$ strictly larger 
than $Z$. From $X'\subseteq \cls{X} \subseteq Z \subset Z'$
and Lemma~\ref{chain}, $\cls{X}$ would be also a
$\gamma$-antecedent of $Z'$, which would contradict
the maximality of $Z$. Therefore,
$X'$ cannot be a $\gamma$-antecedent of a closed set strictly
larger than $Z$ and, together with the facts that define $X'$,
we have that $X'$ is a basic $\gamma$-antecedent of $Z$ whence
$X'\to Z-X'\in\myB$.

We gather the following inequalities: $X'\subseteq\cls{X}$
and $XY \subseteq Z = \cls{Z} = \cls{X'(Z-X')}$;
this is exactly what we need to infer that 
$\B,\{ X'\to Z-X'\}\models X\implies Y$
from Theorem~\ref{closredchar}.\qed

Now we can move to the main result of this section: this basis 
has a minimum number of rules among all bases that are 
complete for the partial rules, according to closure-based 
redundancy with respect to~$\B$.

\begin{thm}
Fix a dataset $\D$, and let $\R$ be the set of rules
that hold with confidence $\gamma$ in $\D$.
Let $\B$ be a basis for the set of implications in $\R$.
Let $\B'\subseteq\R$ be an arbitrary basis, 
having closure-based completeness for $\R$ with
respect to~$\B$.
Then, $\B'$ must have at least as many rules as $\myB$. 
\label{minimality}
\end{thm}

\proof
First, we will prove the following intermediate claim:
for each partial rule in $\myB$, say $X\implies Y-X$, 
there is in $\B'$ a corresponding partial rule of the form
$X'\implies Y'$ with $\cls{X'Y'} = Y$ and 
$\cls{X'} = X$.
We pick any rule $X \implies Y-X\in\myB$, that is, 
where $X$ is a basic $\gamma$-antecedent 
of $Y$; this rule must be redundant, relative
to the implications in $\B$, with respect to 
the new basis $\B'$ under consideration:
for some rule $X'\to Y'\in\B'$, we have that
$\B,\{ X'\to Y' \}\models X\implies Y-X$
which, by Theorem~\ref{closredchar}, is the same as
$X'\subseteq\cls{X}=X$ and $Y \subseteq \cls{X'Y'}$,
together with $c(X'\to Y')\geq\gamma$.
We consider some support ratios:
$\frac{s(\cls{X'Y'})}{s(X)} = \frac{s(X'Y')}{s(\cls{X})} \geq 
\frac{s(X'Y')}{s(X')} \geq \gamma$,
which means that $X$ is a $\gamma$-antecedent
of $\cls{X'Y'}$, a closed set including $Y$;
by the second condition in the definition of
basic $\gamma$-antecedent, this cannot
be the case unless $\cls{X'Y'} = Y$.

Then, again, $c(X'\to Y) = c(X'\to X'Y') = c(X'\to Y')\geq\gamma$,
that is, $X'$ is a $\gamma$-antecedent of $Y$,
and $\cls{X'}\subseteq \cls{Y} = Y$ is as well; 
but $\cls{X'}\subseteq\cls{X}=X$ and,
by minimality of $X$ as a basic $\gamma$-antecedent of $Y$,
it must be that $\cls{X'}=X$. 

Now, to complete the proof of the theorem, we observe that each 
such rule $X' \implies Y'$ in $\B'$ determines
univocally both closed sets $X$ and $Y$, so that the same rule 
in $\B'$ cannot correspond to more than one
of the rules in $\myB$.
This requires $\B'$, therefore, to have at least 
as many rules as $\myB$.\qed

In applications of $\myB$, one needs, in general, 
as a basis both $\myB$ and a basis for the implications, 
such as the GD-basis.
On the other hand, in many practical cases, 
implications provide
little new knowledge, most often just showing existing
(and known) properties of the attributes. If a user 
is satisfied with the $\myB$ basis, and does not ask 
for a basis for the implications nor the representative rules,
then (s)he may get results faster, since in this case the
algorithms would not need to compute minimal 
generators, and just mining closures and their
supports (and organizing them via the subset
relation) would suffice.

Note that the joint consideration of the 
GD-basis and $\myB$ incurs the risk of being
a larger set of rules than the representative rules,
due to the fact that some rules in the GD-basis could be,
in fact, plainly redundant (ignoring the closure-related
issues) with a representative rule. We have observed
empirically that, at high confidence thresholds,
the representative rules tend to be a large basis
due to the lack of specific minimization of implications,
whereas the union of the GD-basis and $\myB$ tends
to be quite smaller; conversely, at lower confidence levels,
the availability of many partial rules increases the
chances of covering a large part of the GD-basis,
so that the representative rules are a smaller basis
than the union of $\myB$ plus GD, even if they are
more in number than $\myB$.
That is: closure-based redundancy may be either 
stronger or weaker, in terms of the optimum basis 
sizes, than plain redundancy. 
Sometimes, $\myB$ even fully coincides with the 
partial representative rules. This is, in fact,
illustrated in the following example.

\begin{exa}
We revisit the example in Figure~\ref{smallex}.
As indicated at the end of Section~\ref{subsecAgYubasis},
the basis for implications consists of six rules:
$AC\dto B$, $AD\dto B$, $BC\dto A$, $BD\dto A$, $CF\dto D$, 
and $DF\dto C$; the iteration-free basis \cite{Wild} 
and the Guigues-Duquenne basis \cite{GD} coincide here,
and these implications are also the representative rules 
at confidence~1.
At confidence $\gamma=0.75$, these
are kept and four {\em representative rules} are added:
$A\to B$, $B\to A$, $AB\to C$, and $D\to C$. 
Since the four left-hand sides are, actually,
closed sets, which is not guaranteed in general,
the basis $\myB$ at this confidence includes
exactly these four rules: no other closure is
a basic $\gamma$-antecedent.

However, if the confidence threshold is lowered to $\gamma = 0.6$, 
we find seven rules in the~$\B^*_{0.6}$ basis:
$A\to BC$, $B\to AC$, $C\to D$, $D\to C$, $CD\to F$, and $F\to CD$,
plus the somewhat peculiar $\emptyset\to C$, since indeed the support
of $C$ is above the same threshold; 
the rules $A\to B$, $B\to A$, 
and $AB\to C$ also hold, but they 
are redundant with respect to $A\to BC$ or $B\to AC$:
$A$ and $B$ are $\gamma$-antecedents of $AB$ but are 
not basic (by way of being also $\gamma$-antecedents of $ABC$), 
whereas $AB$ is a $\gamma$-antecedent of $ABC$ but is not basic 
either since it is not minimal.

Additionally, the sizes of the rules can be reduced somewhat:
$A\to C$ suffices to give $A\to BC$ or indeed $A\to ABC$ since
$A\to C$ is equivalent by reflexivity to $A\to AC$ and there is
a full-confidence implication $AC\dto B$ in the GD-basis that
gives us $A\to ABC$. This form of reasoning is due to \cite{Zaki},
and a similar argument can be made for several of the
other rules. Alternatively, there exists the option of omitting 
those implications that, seen as partial rules, are already 
covered by a partial rule: in this example, these are $AC\dto B$ 
and $BC\dto A$, covered by $A\to BC$ (but {\em not} by $A\to C$,
which needs $AC\dto B$ to infer $A\to BC$); similarly, 
$CF\dto D$ and $CD\dto F$ are plainly redundant with $C\to DF$.
In fact, it can be readily checked that the seven partial
rules in $\B^*_{0.6}$ plus the two remaining implications
in the GD-basis, $AD\dto B$ and $BD\dto A$, form exactly
the representative rules at this confidence threshold.
\end{exa}

\subsection{Double-Support Mining}
\label{suppbound}

For many real-life datasets, including all the standard benchmarks
in the field, the closure space is huge, and reaches easily 
hundreds of thousands of nodes, or indeed even millions.
A standard practice, as explained in the introduction, 
is to impose a support constraint, that is, to ignore (closed) 
sets that do not appear often enough. It has been observed also
that the rules removed by this constraint are often appropriately so,
in that they are less robust and prone to represent statistical
artifacts rather than true information \cite{MeggSrik}.
Hence, we discuss briefly what happens to our basis proposal
if we work under such a support constraint. 

For a dataset $\D$ and confidence and support
thresholds $\gamma$ and $\tau$, respectively,
denote by $\R_{\gamma,\tau}$ the set of rules 
that hold in $\D$ with confidence at least
$\gamma$ and support at least $\tau$. 
We may want to construct either of 
two similar but different sets of rules:
we can ask just how to compute the 
set of rules in $\myB$ that reach that 
support or, more likely, we may
wish a minimum-size basis for 
$\R_{\gamma,\tau}$. We solve both problems.

We first discuss a minimum-size basis for
$\R_{\gamma,\tau}$. Of course, the natural approach
is to compute the rule basis exactly as 
before, but only using closed sets above 
the support threshold. Indeed this works:

\begin{prop}
Fix a dataset $\D$. For any fixed confidence threshold 
$\gamma$ and support threshold $\tau$, 
the construction of basic $\gamma$-antecedents, applied only 
to closed sets of support at least $\tau$, provides a 
minimum-size basis for $\R_{\gamma,\tau}$.
\end{prop}

\proof 
Consider any rule $X\to Y$ of support at least $\tau$
and confidence at least $\gamma$. Then $\cls{X}$ is a 
$\gamma$-antecedent of $\cls{XY}$; also,
$s(\cls{X}) = s(X) 
\geq s(XY) = s(\cls{XY}) \geq\tau$.

Arguing as in the
proof of Theorem~\ref{newsoundcomplete} but restricted
to the closures with support at least $\tau$, we can
find a rule $X'\to Y'-X'$ where both $\cls{X'}$ and 
$\cls{X'Y'}$ have support at least $\tau$, $\cls{X'}$
is a basic $\gamma$-antecedent of $\cls{X'Y'}$, and 
such that $X'\subseteq\cls{X}$ and $XY\subseteq\cls{X'Y'}$
so that it covers $X\to Y$. Minimum size is argued
exactly as in the proof of Theorem~\ref{minimality}:
following the same steps, one proves that
any complete basis consisting of rules in $\R_{\gamma,\tau}$ 
must have separate rules to cover each of the rules 
formed by basic $\gamma$-antecedents of closures of 
support~$\tau$.\qed

We are therefore safe if we apply the basis construction 
for $\myB$ to a lattice of frequent closed sets above 
support $\tau$, instead of the whole lattice of closed sets. 
However, this fact does not ensure that the basis obtained 
coincides with the set of rules in the whole basis $\myB$
having support above $\tau$. There may be rules that are
not in $\myB$ because a large closure, of low support,
prevents some $X$ from being a basic antecedent. If the
large closure is pruned by the support constraint, then $X$
may become a basic antecedent. The following result 
explains with more precision the relationship between
the basis $\myB$ and the rules of support~$\tau$.

\begin{prop}
Fix a dataset $\D$, a confidence threshold 
$\gamma$, and a support threshold~$\tau$.
Assume that $X\subseteq Y$ and that $s(Y)\geq\tau$; 
then $X\to Y-X\in\myB$ 
if and only if $X$ is a basic $\gamma$-antecedent 
of $Y$ in the set of all closures of support
at least $\gamma\times\tau$. 
\end{prop}

This proposition says that, in order to find 
$\myB \cap \R_{\gamma,\tau}$, that is, the set of
rules in $\myB$ that have support at least $\tau$,
we do not need to compute all the closures and
construct the whole of $\myB$; it suffices to
perform the $\myB$ construction on the set
of closures of support $\gamma\times\tau$.
Of course, in both cases we must then 
discard the rules of support less than $\tau$.
We call this sort of process {\em double-support mining}:
given user-defined $\gamma$ and $\tau$, use 
the product to find all closures of support
$\gamma\times\tau$, compute $\myB$ on these
closures, and finally prune out the rules 
with support less than $\tau$ to obtain
$\myB \cap \R_{\gamma,\tau}$, if that is what
is desired.

\proof
Consider a pair of closed sets $X\subset Y$ 
with $s(X)>s(Y)\geq\tau$; we must discuss 
whether $X$ is a basic $\gamma$-antecedent 
of~$Y$ in two different closure lattices: 
the one of all the closed sets and the one 
of frequent closures at support 
threshold~$\gamma\times\tau$. 

The properties of being a $\gamma$-antecedent and of being
minimally so refer to $X$ and $Y$ themselves or to even 
smaller sets, and are therefore unaffected by the support 
constraint. We must discuss just the existence of some
proper superset of $Y$ having $X$ as a $\gamma$-antecedent.
In case $X$ is a basic $\gamma$-antecedent of $Y$, no proper
superset $Z$ of $Y$ has $X$ as $\gamma$-antecedent, whatever 
the support of $Z$; therefore, $X$ will be found to be a basic 
$\gamma$-antecedent of $Y$ also in the smaller lattice of 
frequent closures. 

To show the converse, it suffices to argue that, 
for any proper superset $Z$ of $Y$,
if $X$ is a $\gamma$-antecedent of $Z$,
then $s(Z)\geq\gamma\times\tau$. Indeed,
$
s(Z)\geq \gamma s(X) \geq \gamma\times\tau
$;
hence, if no such
$Z$ is found in the frequent closures lattice 
at support threshold~$\gamma\times\tau$, no such
$Z$ exists at all.\qed

\subsection{Empirical Evaluation}
\label{experim}

Whereas our interests in this paper are rather foundational,
we wish to describe briefly the direct applicability
of our results so far. We have chosen an approach that 
conveniently uses as a black-box a separate closed itemsets 
miner due to Borgelt \cite{BorgeltApriori}. 
We have implemented a construction of the GD basis
using a hypergraph transversal method to construct
representative rules of confidence~1 following the
guidelines of \cite{PT} 
and subsequently
simplifying them to obtain the GD basis
as per \cite{AriasBal}; and we have
implemented a simple algorithm that scans repeatedly
the closed sets mined by the separate program and 
constructs all basic $\gamma$-antecedents. A first
scan picks up $\gamma$-antecedents from the proper
closed subsets and filters them for minimality; once
all minimal antecedents are there for all closures,
a subsequent scan filters out those that are not
basic by way of being antecedents of larger sets.
Effectively the algorithm does not implement the
definition but the immediate extension of the 
characterization in Corollary~\ref{charreprbasisAY}
to the closure-based case.

A natural alternative consists in preprocessing the lattice
as a graph in order to find the predecessors of a node directly; 
however, in practice, with this alternative, whenever the graph 
requires too much space, we found that the computation slows down 
unacceptably, probably due to a worse fit to virtual memory caching. 
Our implementation gives us answers in just seconds in most cases, 
on a mid-range Windows XP laptop, taking a few minutes when the closure 
space reaches a couple dozen thousand itemsets.

\begin{table}[t]
\begin{center}
\begin{tabular}{|l||r|r|r|r|r|r|r|r|}
	\hline
Dataset & S/C & Traditional & Closure-based & RR Imp & GD & $\myB$ & Sum \\
	\hline \hline
 Chess  & 80 & 552564 & 27711 & 2228 & 5 & 226 & 231 \\
 Chess  & 70 & 8171198 & 152074 & 13193 & 10 & 891 & 901  \\
 Connect  & 97 & 8092 & 1116 & 161 & 4 & 41 & 45  \\
 Connect  & 90 & 3640704 & 18848 & 3359 & 14 & 222 & 236 \\
 Mushroom  & 40 & 7020 & 475 & 170 & 24 & 41 & 65 \\
 Mushroom  & 20 & 19191656 & 5741 & 1739 & 177 & 158 & 328 \\
 Pumsb   & 95 & 1170    & 267 & 62 & 2 & 32 & 34\\
 Pumsb   & 85 & 1408950 & 44483 & 9559 & 9 & 1080 & 1089\\
 Pumsb\_star & 60  & 2358 & 192 & 43 & 5 & 6 & 11\\
 Pumsb\_star & 40 & 5659536 & 13479 & 2939 & 48 & 82 & 129 \\
 T10I4D100K & 0.5 & 2216 & 1231 & 0 & 0 & 585 & 585 \\
 T10I4D100K & 0.1 & 431838 & 86902 & 582 & 214 & 4054 & 4268 \\
	\hline
\end{tabular}
\smallskip
\caption{Number of rules in various bases for benchmark datasets.}
\label{maintable}
\end{center}
\end{table}

On the basis of this implementation, we have undertaken some
empirical evaluations of the sizes of the basis. We consider
that the key point of our contribution is the mathematical
proof of absolute size minimality, but, as a mere illustration, 
we show the figures of some of the cases explored in \cite{Zaki}
in Table~\ref{maintable}. 
The datasets and thresholds are set exactly as per that reference;
column ``S/C" is the confidence and support parameters.
Columns ``Traditional'' (for the number of rules under
the standard traditional definition \cite{AIS}) 
and ``Closure-based'' (for the number
of rules obtained by the closure-based method proposed 
in \cite{Zaki}) are
taken verbatim from the same reference. We have added the number
of rules in the representative basis for implications 
at 100\% confidence ``RRImp'', that coincides with
the iteration-free basis \cite{Wild} and other proposals
as discussed 
at the beginning of Subsection~\ref{subscharclosred}; 
the size of the GD basis for the 
same implications (often yielding huge savings);
and the number of rules in the $\myB$ basis of 
partial rules, which, in the totality of these cases,
did coincide with the representative rules at the
corresponding thresholds. As discussed in
the end of Section~\ref{mybasissect}, 
representative rules encompass implications but
$\myB$ must be taken jointly with the GD basis,
so we give also the corresponding sum.

\begin{figure}[t]
\begin{center}
\includegraphics[width=14cm]{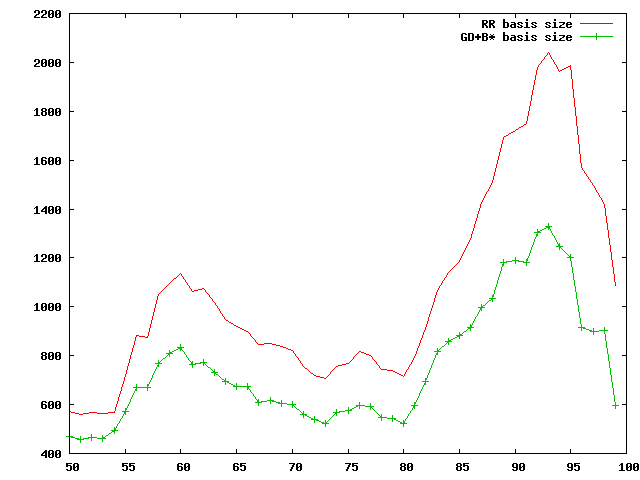}
\end{center}
\caption{Basis sizes per confidence in
{\tt pumsb-star} at 40\% support}
\label{pumsbstar40}
\end{figure}

\begin{figure}[t]
\begin{center}
\includegraphics[width=14cm]{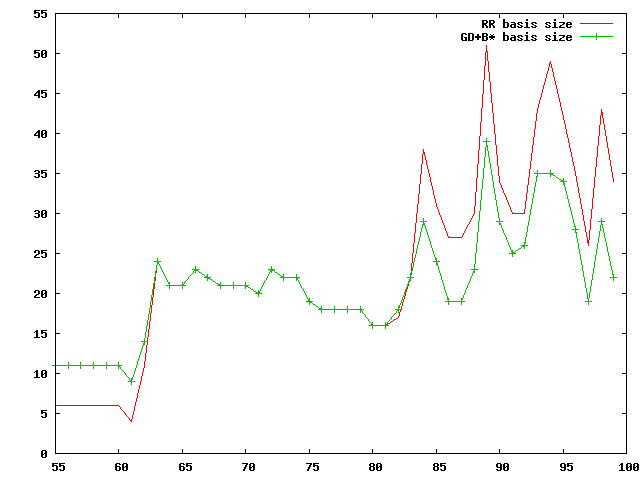}
\end{center}
\caption{Basis sizes per confidence in
{\tt pumsb-star} at 60\% support}
\label{pumsbstar60}
\end{figure}

The confidence chosen in \cite{Zaki} for this comparison,
namely, coincident with the support threshold, is, in our 
opinion, too low to provide a good perspective; at these
thresholds, representative rules essentially correspond 
to support bounds (rules with empty left-hand~side).
To complement the intuition, we provide the evolution
of the sizes of the representative rules and the $\myB$
basis for the dataset {\tt pumsb-star}, downloaded from \cite{FIMI},
at the same support thresholds of 40\% and 60\% used in 
Table~\ref{maintable}, with confidence ranging from 99\% to 51\%,
at 1\% granularity. 
The Guigues-Duquenne bases at these support thresholds 
consist of~48 and~5 rules respectively. 
These have been added to the size of $\myB$
in Figures \ref{pumsbstar40}~and~\ref{pumsbstar60}.
At these confidence tresholds,
the traditional notion of association rules gives 
from 105086 up to 179684 rules at support 40\%,
and
between 268 and 570 rules at support~60\%.
Note that, in that notion, association rules are restricted,
by definition, to singleton consequents; larger numbers
would be found if this condition is lifted for a fairer
comparison with the bases we study.
These figures show the advantage of the closure-based basis
over representative rules up to the point where the implications 
become subsumed by partial representative rules.

We want to point out as well one interesting aspect
of the figures obtained.
The standard settings for association rules lead to a monotonicity 
property, by which lower confidence thresholds allow for more rules, 
so that the size of the output grows (sometimes enormously) as the 
confidence threshold decreases. However, in the case of the $\myB$
basis and the representative rules,
some datasets exhibit a nonmonotonic evolution: at lesser confidence
thresholds, sometimes less rules are obtained. Inspecting the actual 
rules, we can find the reason: sometimes there are several rules at, 
say, 90\% confidence that become simultaneously redundant due to a 
single rule of smaller confidence, say 85\%, which does not appear 
at 90\% confidence. This may reduce the set of rules upon lowering 
the confidence threshold. 

\section{Towards General Entailment}
\label{closbasedent}

We move on towards a further contribution of this paper: we propose 
a stronger notion of redundancy, as progress towards a complete logical 
approach, where redundancy would play the role of entailment and a
sound and complete deductive calculus is sought. Considering the 
redundancy notions described so far, the following question naturally 
arises: beyond all these notions of redundancy that relate one partial 
rule to another partial rule, possibly in presence of 
implications, is it indeed possible that a partial rule is entailed 
jointly by two partial rules, but not by a single one of them? and,
if so, when does this happen? We will fully answer this question below.

The failures of Transitivity and Augmentation may suggest the 
intuition of a negative answer: it looks like any combination 
of two partial rules of confidence at least $\gamma$, but with 
$\gamma<1$, will require us to 
multiply confidences, 
reaching as low as $\gamma^2$ or lower; but this intuition is 
wrong. We will characterize precisely the case where, at a fixed
confidence threshold, a partial 
rule follows from exactly two partial rules, a case where our 
previous calculus becomes incomplete; and we will identify one 
extra deduction scheme that allows us to conclude as consequent 
a partial rule from two premise partial rules in a sound form.
The calculus obtained is complete with respect to entailment 
from two premise rules. We present the whole setting in terms 
of closure-based redundancy, but the development carries over 
for plain redundancy, simply by taking the identity as closure
operator.

A first consideration is that we no longer have a single value of the 
confidence to compare; therefore, we take a position like the one in 
most cases of applications of association rule mining in practice, 
namely: we fix a confidence threshold, and consider only rules whose 
confidence is above it. An alternative view, further removed from 
practice,
would be to require just that the confidence of all our 
conclusions should be at least the same as the minimum of the 
confidences of the premises.

As an example, consider the following fact (the analogous statement 
for $\gamma<1/2$ does not hold, as discussed below):

\begin{prop}
Let $\gamma\geq1/2$.
Assume that items $A$, $B$, $C$, $D$ are present in~$\U$ and that
the confidence of the rules $A\to BC$ and $A\to BD$ is above
$\gamma$ in dataset~$\D$. Then, the confidence of the rule
$ACD\to B$ in $\D$ is also above $\gamma$.
\label{example}
\end{prop}

We do not provide a formal proof of this claim
since it is just the simplest particular case of 
Theorem~\ref{dospremisas} below. We consider the 
following definition:

\begin{defi}
Given a set $\B$ of implications, and a set $\R$ of partial rules,
rule $X_0\to Y_0$ is $\gamma$-redundant with respect to them
(or also $\gamma$-entailed by them), denoted
$\B,\R \models_{\gamma} X_0\to Y_0$,
if every dataset in which the rules of $\B$ have confidence~1
and the confidence of all the rules in $\R$ is at least $\gamma$
must satisfy as well $X_0\to Y_0$ with confidence at least $\gamma$.
The entailment is called ``proper'' if it does not hold for 
proper subsets of $\R$; otherwise it is ``improper''.
\label{fullredundancy}
\end{defi}

Note that, in this case, the parameter $\gamma$ is necessary to
qualify the entailment relation itself. In previous sections
we had a mere confidence inequality that did not depend on $\gamma$.
The main result of this section is now:

\begin{thm}
Let $\B$ be a set of implications, and let $1/2 \leq\gamma<1$. 
Consider three partial rules, $X_0\to Y_0$, $X_1\implies Y_1$, 
and $X_2\implies Y_2$.
Then, 
$\B, \{X_1\implies Y_1, \, X_2\implies Y_2\} \models_{\gamma} X_0\to Y_0$
if and only if either:
\begin{enumerate}[\em(1)]
\item $Y_0\subseteq \cls{X_0}$, or
\item $\B,\{X_1\to Y_1\} \models X_0\to Y_0$, or
\item $\B,\{X_2\to Y_2\} \models X_0\to Y_0$, or
\item all the following conditions simultaneously hold:
\smallskip
\begin{enumerate}[\em(i)]
\item $X_1 \subseteq \cls{X_0}$ 
\item $X_2 \subseteq \cls{X_0}$ 
\item $X_1 \subseteq \cls{X_2Y_2}$ 
\item $X_2 \subseteq \cls{X_1Y_1}$ 
\item $X_0 \subseteq \cls{X_1Y_1X_2Y_2}$
\item $Y_0 \subseteq \cls{X_0Y_1}$
\item $Y_0 \subseteq \cls{X_0Y_2}$
\end{enumerate}
\end{enumerate}
\label{dospremisas}
\end{thm}

\proof 
Let us discuss first the leftwards implication.
In case (1), rule $X_0\to Y_0$ holds trivially.
Clearly cases (2) and (3) also give (improper) entailment.
For case (4),
we must argue that, if all the seven conditions hold, then
the entailment relationship also holds. 
Thus, fix any dataset~$\D$ where
the confidences of the premise rules are at least $\gamma$:
these assumptions can be written, respectively, 
$
  s(X_1Y_1)\geq\gamma s(X_1)
$
and
$
  s(X_2Y_2)\geq\gamma s(X_2)
$, or equivalently for the corresponding closures.

We have to show that the confidence of $X_0\to Y_0$ in $\D$
is also at least $\gamma$. Consider the following four sets 
of transactions from $\D$:

$
  A = \{ t \in\D\st t \models X_0Y_0 \}
$

$
  B = \{ t \in\D\st t \models X_0, t \not\models X_0Y_0 \}
$

$
  C = \{ t \in\D\st t \models X_1Y_1, t \not\models X_0 \}
$

$
  D = \{ t \in\D\st t \models X_2Y_2, t \not\models X_0 \}
$

\noindent
and let $a$, $b$, $c$, and $d$ be the respective cardinalities.

We first argue that all four sets are mutually disjoint.

 This is easy for most pairs: clearly $A$ and $B$ have
 incompatible behavior with respect to $Y_0$; and a tuple
 in either $A$ or $B$ has to satisfy $X_0$, which makes it
 impossible that that tuple is accounted for in either $C$
 or $D$. The only place where we have to argue a bit more 
 carefully is to see that $C$ and $D$ are disjoint as well:
 but a tuple $t$ that satisfies both $X_1Y_1$ and $X_2Y_2$,
 that is, satisfies their union $X_1Y_1X_2Y_2$, must satisfy
 every subset of the corresponding closure as well, such as
 $X_0$, due to condition (v). Hence, $C$ and $D$ are disjoint.

Now we bound the supports of the involved itemsets as follows:
clearly, by definition of $A$, $s(X_0Y_0) = a$. All tuples that
satisfy $X_0$ are accounted for either as satisfying $Y_0$ as well,
in $A$, or in $B$ in case they don't; disjointness then guarantees
that $s(X_0) = a+b$.

We see also that $s(X_1)\geq a+b+c+d$, because $X_1$ is satisfied
by the tuples in $C$, by definition; by the tuples in $A$ or $B$,
by condition (i); and by the tuples in $D$, by condition (iii);
again disjointness allows us to sum all four cardinalities.
Similarly, using instead (ii) and (iv), we obtain $s(X_2)\geq a+b+c+d$.

 The next delicate point is to show an upper bound on $s(X_1Y_1)$ 
(and on $s(X_2Y_2)$ symmetrically).
We split all the tuples that satisfy $X_1Y_1$ into
two sets, those that additionally satisfy $X_0$, and those that don't.
 Tuples that satisfy $X_1Y_1$ and not $X_0$ are exactly those in $C$,
 and there are exactly $c$ many of them. Satisfying $X_1Y_1$ and $X_0$ 
 is the same as satisfying $X_0Y_1$ by condition (i), and tuples that do 
 it must also satisfy $Y_0$ by condition (vi). Therefore, they satisfy
 both $X_0$ and $Y_0$, must belong to $A$, and there can be at most
 $a$ many of them. That is, 
$s(X_1Y_1) \leq a+c$ and, symmetrically, 
resorting to (ii) and (vii), $s(X_2Y_2)\leq a+d$.

Thus we can write the following inequations:
$$
        a+c \geq s(X_1Y_1) \geq \gamma s(X_1) \geq \gamma(a+b+c+d)
$$
$$
        a+d \geq s(X_2Y_2) \geq \gamma s(X_2) \geq \gamma(a+b+c+d)
$$
Adding them up, 
using $\gamma\geq\frac{1}{2}$, 
 we get
 $$
         2a+c+d \geq 2\gamma(a+b+c+d) 
                = 2\gamma(a+b) + 2\gamma(c+d) \geq 2\gamma(a+b)+c+d
 $$
 that is, 
$a\geq\gamma(a+b)$, so that
 $$
        c(X_0\to Y_0) = \frac{s(X_0Y_0)}{s(X_0)} = \frac{a}{a+b} \geq\gamma
 $$
as was to be shown.

 Now we prove the rightwards direction;
the bound $\gamma\geq\frac{1}{2}$ is not necessary 
 for this part. 
Since all our supports are integers,
 we can assume that the threshold is a rational number, 
 $\gamma=\frac{m}{n}$, so that
 we can count on $n-m>0$ and $1\leq m\leq n-1$.
We will argue the contrapositive, assuming 
that we are in neither of the
four cases, and showing that the
entailment does not happen, that is, 
it is possible to construct
a counterexample dataset for which 
all the implications in $\B$ hold,
and the two premise partial rules 
have confidence at least~$\gamma$, 
whereas the rule in the conclusion
has confidence strictly below~$\gamma$. 
This requires us to construct
a number of counterexamples 
through a somewhat long case 
analysis. 
In all of them, all the tuples will be closed sets
with respect to $\B$; this ensures that these implications are
satisfied in all the transactions. 
We therefore assume that case (1) does not happen, 
that is, $Y_0\not\subseteq \cls{X_0}$; and that 
cases (2) and (3) do not happen
either. Now, Theorem~\ref{closredchar} tells us
that 
$X_1\subseteq \cls{X_0}$ implies $X_0Y_0 \not\subseteq \cls{X_1Y_1}$,
and that
$X_2\subseteq \cls{X_0}$ implies $X_0Y_0 \not\subseteq \cls{X_2Y_2}$.
Along the rest of the proof, we will refer to 
the properties explained in this
paragraph as the ``known facts''.

Then, assuming that case (4) does not hold either, we have to consider
multiple ways for the conditions (i) to (vii) to fail. Failures of
(i) and (ii), however, cannot be argued separately, and we discuss
them together.

\smallskip
\noindent{\sl Case A}. Exactly one of (i) and (ii) fails. By symmetry,
renaming $X_1\to Y_1$ into $X_2\to Y_2$ if necessary, we can assume that
(i) fails and (ii) holds. Thus, $X_1 \not\subseteq \cls{X_0}$ but 
$X_2 \subseteq \cls{X_0}$. 
Then, by the known facts, $X_0Y_0 \not\subseteq \cls{X_2Y_2}$.
We consider a dataset consisting of one transaction with the 
itemset $\cls{X_2Y_2}$, $mn-1$ transactions with the set 
$\cls{X_0X_1Y_1X_2Y_2}$,
and $n(n-m)$ transactions with the set $\cls{X_0}$, for a total of $n^2$ 
transactions. Then, the support of $X_0$ is either $n^2-1$ or $n^2$, and
the support of $X_0Y_0$ is at most $mn-1$, for a confidence bounded
by $\frac{mn-1}{n^2-1} < \frac{mn}{n^2} = \gamma$ for the rule $X_0\to Y_0$.
However, the premise rules hold: since (i) fails, the support of 
$X_1$ is at most $mn$, and the support of $X_1Y_1$ is at least $mn-1$,
for a confidence at least $\frac{mn-1}{mn} \geq \frac{m}{n} = \gamma$ 
for $X_1\to Y_1$;
whereas the support of $X_2$ is $n^2$, that of $X_2Y_2$ is at least $nm$, and
therefore the confidence is at least $m/n = \gamma$.

\smallskip
\noindent{\sl Case B}. This corresponds to both of (i) and (ii) failing.
Then, for a dataset consisting only of $\cls{X_0}$'s, the premise rules
hold vacuously whereas $X_0\to Y_0$ fails. We can also avoid
arguing through rules holding vacuously by means of a dataset consisting
of one transaction $\cls{X_0X_1Y_1X_2Y_2}$ and 
$n$ transactions~$\cls{X_0}$.

\smallskip
\noindent{\sl Remark}. 
For the rest of the cases, we will assume that both of (i) and (ii) hold,
since the other situations are already covered. Then, by the known facts,
we can freely use the properties $X_0Y_0 \not\subseteq \cls{X_1Y_1}$ and
$X_0Y_0 \not\subseteq \cls{X_2Y_2}$.

\smallskip
\noindent{\sl Case C}. Assume (iii) fails, $X_1 \not\subseteq \cls{X_2Y_2}$,
and consider a dataset consisting of one transaction $\cls{X_0}$, 
$n$ transactions
$\cls{X_1Y_1}$, and $n^2$ transactions $\cls{X_2Y_2}$. 
Here, by the known facts, the
support of $X_0Y_0$ is zero. It suffices to check that the antecedent rules
hold. Since (iii) fails, and (i) holds, the support of $X_1$ is 
exactly $n+1$ and the 
support of $X_1Y_1$ is at least $n$, for a confidence of at least 
$\frac{n}{n+1} > \frac{n-1}{n} \geq \frac{m}{n} = \gamma$; whereas the 
support of $X_2$ is at most $n^2 + n + 1$ (depending on whether (iv) holds)
for a confidence of rule $X_2\to Y_2$ of at least $\frac{n^2}{n^2+n+1}$ which 
is easily seen to be
above $\frac{n-1}{n} \geq \frac{m}{n} = \gamma$.
 
The case where (iv) fails is fully symmetrical and can be argued just 
interchanging the roles of $X_1\to Y_1$ and $X_2\to Y_2$.

\smallskip
\noindent{\sl Case D}. Assume (v) fails. It suffices to consider a dataset
with one transaction $\cls{X_0}$ and $n-1$ 
transactions $\cls{X_1Y_1X_2Y_2}$. Using (i) 
and (ii), for both premises the confidence is $\frac{n-1}{n}\geq\gamma$,
the support of $X_0$ is 1, and the support of $X_0Y_0$ is zero by
the known fact $Y_0\not\subseteq \cls{X_0}$ and the failure of (v).

\smallskip
\noindent{\sl Case E}. We assume that (vi) fails, but a symmetric argument
takes care of the case where (vii) fails. Thus, we have 
$Y_0 \not\subseteq \cls{X_0Y_1}$. By treating this case last, we can
assume (i), (ii), and (v) hold, and also the known facts that
$X_0Y_0 \not\subseteq \cls{X_1Y_1}$ and $X_0Y_0 \not\subseteq \cls{X_2Y_2}$.
We consider a dataset with one 
transaction $\cls{X_0Y_1}$, one transaction $\cls{X_2Y_2}$, 
$m-1$ transactions
$\cls{X_1Y_1X_2Y_2}$, and $n-m-1$ transactions~$\cls{X_0}$ 
(note that this last
part may be empty, but $n-m-1\geq0$; the total is $n$ transactions).
By (v), the support of $X_0$ is at least $n-1$, whereas the support
of $X_0Y_0$ is at most $m-1$, given the available facts. Since
$\frac{m-1}{n-1} < \gamma$, rule $X_0\to Y_0$ does not hold. However,
the premises hold: all supports are at most $n$, the total size, 
and the supports of $X_1Y_1$ (using (i)) and $X_2Y_2$ are both $m$.

This completes the proof.\qed

A small point that remains to be clarified is the role of the
condition \hbox{$\gamma\geq1/2$}. As indicated in the proof of
the theorem, that condition is only necessary in one of the
two directions. If there is entailment, the conditions enumerated
must hold irrespective of the value of $\gamma$. In fact, for 
$0<\gamma<1/2$, proper entailment from a set 
of two (or more) premises never holds, and $\gamma$-entailment 
in general is characterized as (closure-based) redundancy as 
per Theorem~\ref{closredchar} and the corresponding calculus.
Indeed:

\begin{thm}
Let $0<\gamma<1/2$. 
Then, 
$\B, \{X_1\implies Y_1, \, X_2\implies Y_2\} \models_{\gamma} X_0\to Y_0$
if and only if either:
\begin{enumerate}[\em(1)]
\item $Y_0\subseteq \cls{X_0}$, or
\item $\B,\{X_1\to Y_1\} \models X_0\to Y_0$, or
\item $\B,\{X_2\to Y_2\} \models X_0\to Y_0$.
\end{enumerate}
\label{menorunmedio}
\end{thm}

\proof 
The leftwards proof is already part of 
Theorem~\ref{dospremisas}. For the converse, assume
that the three conditions fail: similarly to the
previous proof, we have as known facts the following:
$Y_0\not\subseteq \cls{X_0}$,
$X_1\subseteq \cls{X_0}$ implies $X_0Y_0 \not\subseteq \cls{X_1Y_1}$
and
$X_2\subseteq \cls{X_0}$ implies $X_0Y_0 \not\subseteq \cls{X_2Y_2}$.
We prove that there are datasets giving low confidence
to $X_0\to Y_0$ and high confidence to both premise rules.
If both $X_1\not\subseteq \cls{X_0}$ and $X_2\not\subseteq \cls{X_0}$ 
then we consider one transaction $\cls{X_1Y_1}$, one transaction 
$\cls{X_2Y_2}$, and a large number $m$ of transactions $\cls{X_0}$
which do not change the confidences of the premises but
lead to a confidence of at most $2/m$ for $X_0\to Y_0$.
Also, if $X_1\not\subseteq \cls{X_0}$ but $X_2\subseteq \cls{X_0}$,
where the symmetric case is handled analogously, we are
exactly as in Case~A in the proof of Theorem~\ref{dospremisas}
and argue in exactly the same way.

The interesting case is when 
both $X_1\subseteq \cls{X_0}$ and $X_2\subseteq \cls{X_0}$;
then
both
$X_0Y_0 \not\subseteq \cls{X_1Y_1}$
and
$X_0Y_0 \not\subseteq \cls{X_2Y_2}$.
We fix any integer $k\geq\frac{\gamma}{1-2\gamma}$ and use 
the fact that $\gamma<1/2$ to ensure that the fraction is
positive and that the inequality can be transformed, 
by solving for $\gamma$,
into $\frac{k}{2k+1}\geq\gamma$ (following these steps
for $\gamma\geq1/2$ either makes the denominator null or 
reverses the inequality due to a negative sign).
We consider a dataset with one transaction for $\cls{X_0}$
and $k$ transactions for each of $\cls{X_1Y_1}$ and $\cls{X_2Y_2}$.
Even in the worst case that either or both of $X_1$ and $X_2$ 
show up in all transactions, the confidences of $X_1\to Y_1$
and $X_2\to Y_2$ are at least $\frac{k}{2k+1}\geq\gamma$,
whereas the confidence of $X_0\to Y_0$ is zero.\qed

\subsection{Extending the calculus}

We work now towards a rule form, in order to enlarge our calculus
with entailment from larger sets of premises. We propose the 
following additional rule:

\smallskip
(2A)\quad
$\frac{\strut
X_1\to Y_1, 
\quad 
X_2\to Y_2, 
\quad 
X_1Y_1\dto X_2, 
\quad 
X_2Y_2\dto X_1, 
\quad 
X_1Y_1X_2Y_2\dto Z_1,
\quad
X_1Y_1Z_1\dto Z_2,
\quad
X_2Y_2Z_1\dto Z_2
}
{\strut
X_1X_2Z_1\to Z_2 
}$

\smallskip
\noindent
and state the following properties:

\begin{thm}
Given a threshold $\gamma\geq1/2$ and a set $\B$ of implications,
\begin{enumerate}[\em(1)]
\item
this deduction scheme is sound, and
\item
together with the deduction schemes 
in Section~\ref{redundcalculus},
it gives a calculus complete with 
respect to all entailments with two 
partial rules in the antecedent.
\end{enumerate}
\label{newrule}
\end{thm}

\proof
This follows easily from Theorem~\ref{dospremisas}, in that 
it implements the conditions of case (4); soundness is seen by 
directly checking that the conditions (i) to (vii) in case 4 of 
Theorem~\ref{dospremisas} hold: let $X_0=X_1X_2Z_1$ and $Y_0=Z_2$;
then, conditions (i) and (ii) hold trivially, and the rest are
explicitly required in the form of implications in the premises
(notice that $X_1Y_1\dto X_2$ implies that $X_1Y_1Z_1\dto Z_2$ 
and $X_1X_2Y_1Z_1\dto Z_2$ are equivalent).
Completeness is argued by considering any rule $X_0\to Y_0$ 
entailed by $X_1\to Y_1$ and $X_2\to Y_2$ jointly with respect 
to confidence threshold $\gamma$; if the entailment is improper, 
apply Theorem~\ref{calculussoundcomplete}, otherwise just apply 
this new deduction scheme with $Z_1=\cls{X_0}$ 
and $Z_2=Y_0$ to get $\cls{X_0}\to Y_0$
and apply $(\ell I)$ to obtain $X_0\to Y_0$.
It is easy to see that the scheme is indeed applicable:
proper entailment implies that all seven conditions in
case (4) hold and, for $Z_1=\cls{X_0}$, we get from (i)
and (ii) that $X_1X_2Z_1 = Z_1$; under this equality,
the remaining five conditions provide exactly the premises
of the new deduction scheme.\qed

\section{Discussion}

Our main contribution, at a glance, is a study of
confidence-bounded association rules in terms of a family of notions 
of redundancy. We have provided characterizations of several existing
redundancy notions; we have described how these previous proposals,
once the relationship to the most robust definitions has been clarified,
provide a sound and complete deductive calculus for each of them; and 
we have been able to prove global optimality of an existing basis 
proposal, for the plain notion of redundancy, and also to improve the 
constructions of bases for closure-based redundancy, up to 
global optimality as well. 

Many existing 
notions of redundancy discuss redundancy of a partial rule only with 
respect to another single partial rule; in our Section~\ref{closbasedent},
we have moved beyond into the use of two partial rules. For this approach 
to redundancy, we believe that this last step has been undertaken for the 
first time here; the only other reference we are aware of, where a
consideration is made of several partial rules entailing a partial rule, 
is the early \cite{Lux}, which used a much more demanding notion of 
redundancy in which the exact values of the confidence of the rules
were both available on the premises and required in the conclusion.
In our simpler context, we have shown that the following holds:
for $0<\gamma<1/2$, there is no case of proper
$\gamma$-entailment from two premises; 
beyond $1/2$,
there are such cases, and they are fully captured
in terms of set inclusion relationships between
the itemsets involved. 
We conjecture that a more general pattern holds.

More precisely, we conjecture the following:
for values of the confidence parameter $\gamma\neq0$, 
such that $\frac{n-1}{n}\leq\gamma<\frac{n}{n+1}$
(where $n\geq1$), there are partial rules that are 
properly entailed from $n$ premises, partial rules
themselves, but there are no proper entailments
from $n+1$ or more premises. That is, intuitively, 
higher values of the confidence threshold correspond, 
successively, to the ability of using more and more 
partial premises.
However, the combinatorics to fully characterize the
case of two premises are already difficult enough 
for the current state of the art, and progress towards
proving this conjecture requires to build intuition to much
further a degree.

This may be, in fact, a way towards stronger 
redundancy notions and always smaller bases 
of association rules. We wish to be able to 
establish such more general methods to reach
absolutely minimum-size bases with respect 
to general entailment, possibly depending on 
the value of the confidence threshold $\gamma$ 
as per our conjecture as just stated.

We observe the following: after constructing
a basis, be it either the representative rules or 
the $\myB$ family, it is a simple matter to scan it
and check for the existence of pairs of rules 
that generate a third rule in the basis according to 
Theorem~\ref{dospremisas}:
then, removing such third rules gives a 
smaller basis with respect to this more 
general entailment. However, we must say 
that some preliminary empirical tests 
suggest that this sort of entailments from 
two premises seems to appear in practice 
very infrequently, so that the check is 
computationally somewhat expensive compared 
to the scarce savings it provides for the
basis~size.

Now that all our contributions are in place,
let us review briefly a point that we made in
the Introduction regarding what is expected
to be the role of the basis.
The statement that association rule mining
produces huge outputs, and that this is indeed
a problem, not only is acknowledged in many 
papers but also becomes self-evident to 
anyone who has looked at the output of any
of the association miner implementations
freely accessible on the web
(say \cite{BorgeltApriori} for one).
However, we do not agree that it is {\em one}
problem: to us, it is, in fact, {\em two} 
slightly different problems, and confusing
them may lead to controversies that are
easier to settle if we understand that 
different persons may be interested in
different problems, even if they are
stated similarly. Specifically, let us
ask whether a huge output of an 
association miner is a problem for the
user, who needs to receive the output
of the mining process in a form that
a human can afford to read and understand,
or for the software that is to store all
these rules, with their supports and
confidences. Of course, the answer is
``both'', but the solutions may not
coincide. 

Indeed, sophisticated conceptual advances
have provided data structures to be computed
from the given dataset in such a way that, 
within reasonable computational 
resource limits, they are able to give
us the support and confidence of any
given rule in the given dataset; maybe
a good approximation is satisfactory 
enough, and this may allow us to obtain
some efficiency advantages.
The set of frequent sets, the set
of frequent closures, and many other
methods have been proposed for this task;
see 
\cite{AMSTV},
\cite{BBR}, 
\cite{GoeSets}, 
\cite{CrisSim}, 
\cite{Lux}, 
\cite{MuTo},
\cite{PasBas}, 
\cite{Zaki}, 
and the surveys \cite{CRB} and \cite{Krysz}.

Our approach is, rather, logical in nature, and
aimed at the other variant
of the problem: what rules are irredundant, in a
general sense.
From these, redundant rules reaching the
thresholds can be found, ``just as rules''.
So, we formalize a situation closer to the 
practitioner's process, where a confidence 
threshold $\gamma$ is enforced beforehand 
and the rules with confidence at least $\gamma$ 
are to be discussed;
but we do {\em not} need to infer from the basis
the value of the confidence of each of these other rules, 
because we can recompute it immediately as a quotient
of two supports, found in an additional data
structure that we assume kept, such as
the closures lattice with the supports of
each closed set.

Therefore, our bases, namely, the already-known 
representative rules and our new closure-based 
proposal $\myB$, are rather ``user-oriented'': 
we know that all rules above the threshold can be obtained
from the basis, and we know how to infer them when
necessary; thus, we could, conceivably, 
guide (or be guided by) the user 
if (s)he wishes to see all the 
rules that can be derived from
one of the rules in the basis; 
this user-guided exploration of 
the rules resulting from the mining
process is alike to the ``direction-setting rules''
of \cite{LiuHsuMa}, with the difference that their
proposal is based on statistical considerations
rather than the logic-based approach we have followed. 

The advantage is that our basis is not required
to provide as much information as the bases we
have mentioned so far, because the notion of
redundancy does not require us to be able to
compute the confidence of the redundant rules.
This is why we can reach an optimum size,
and indeed, compared to \cite{PasBas}~or~\cite{Zaki},
$\myB$ differs because these proposals,
essentially, pick all minimal generators 
of each antecedent, which we avoid.
The difference is marginal in the conceptual sense; however
the figures in practical cases may differ considerably, and
the main advantage of our construction is that we can actually
prove that there is no better alternative as a basis for
the partial rules with respect to closure-based 
redundancy.

Further research may proceed along several questions.
We believe that 
a major breakthrough in intuition is necessary to
fully understand entailment among partial rules in
its full generality, either as per our conjecture
above or against it; variations of our definition
may be worth study as well, such as removing the
separate confidence parameter and requiring that
the conclusion holds with a confidence at least
equal to the minimum of the confidences of the
premises.

Other questions are how to extend this approach to the mining 
of more complex dependencies \cite{SimPurity}
or of dependencies among structured objects; however, 
extending the development to sequences, partial orders, and trees, 
is not fully trivial, because, as demonstrated in~\cite{BBL08},
there are settings where
the combinatorial structures may make redundant 
certain rules that would
not be redundant in a propositional (item-based) framework;
additionally, an intriguing question is: what part of all 
this discussion remains true if implication
intensity measures different from confidence 
(\cite{Garriga}, \cite{GH}) are used?

\section*{Acknowledgements}

The author is grateful to his research group at UPC and to the
regular seminar attendees; also, for many observations, suggestions,
comments, references, and improvements, the author gratefully
acknowledges 
Cristina T\^{\i}rn\u{a}uc\u{a},
Ver\'onica Dahl, 
Tijl de Bie,
Jean-Fran\c{c}ois Boulicaut, 
the participants in seminars where the author has presented this work,
the reviewers of the conference papers where most of the results
of this paper were announced, 
and the reviewers of the present paper.

\bibliographystyle{plain}

\end{document}